\documentclass[10pt,aps,prd,floatfix,twocolumn,preprintnumbers,
superscriptaddress,nofootinbib,amssymb,notitlepage,eqsecnum]{revtex4-2}
\usepackage{bm}
\usepackage[dvipsnames]{xcolor}
\usepackage{amsmath,amsthm,amssymb}
\usepackage{amsfonts}
\usepackage{hyperref}
\hypersetup{hidelinks,
 pdftitle={Observational constraints on scalar-vector-tensor dark energy with phantom-divide crossing},
 pdfauthor={Nandan Roy, Shinji Tsujikawa, Ying-li Zhang, and Zejun Zhang}}
\usepackage{graphicx}
\usepackage[normalem]{ulem}

\allowdisplaybreaks[1]

\newcommand{\rd}{{\rm d}}
\newcommand{\be}{\begin{equation}}
\newcommand{\ee}{\end{equation}}
\newcommand{\ba}{\begin{eqnarray}}
\newcommand{\ea}{\end{eqnarray}}
\newcommand{\Mpl}{M_{\rm Pl}}

\begin{document}

\title{Observational constraints on scalar-vector-tensor dark energy
\texorpdfstring{\\}{ }with phantom-divide crossing}

\author{Nandan Roy}
\email{nandan.roy@mahidol.ac.th}
\affiliation{NAS, Centre for Theoretical Physics \& Natural Philosophy, Mahidol University, Nakhonsawan Campus, Phayuha Khiri, Nakhonsawan 60130, Thailand}

\author{Shinji Tsujikawa}
\email{tsujikawa@waseda.jp}
\affiliation{Department of Physics, Waseda University,
3-4-1 Okubo, Shinjuku, Tokyo 169-8555, Japan}

\author{Ying-li Zhang}
\email{yingli@tongji.edu.cn}
\affiliation{School of Physics Science and Engineering, Tongji University,
Shanghai 200092, China}
\affiliation{Institute for Advanced Study of Tongji University, Shanghai 200092, China}
\affiliation{Asia Pacific Center for Theoretical Physics, Pohang 37673, Korea}
\affiliation{Kavli Institute for the Physics and Mathematics of the Universe (WPI), The University of Tokyo Institutes for Advanced Study, The University of Tokyo, Chiba 277-8583, Japan}

\author{Zejun Zhang}
\email{zjzhang@tongji.edu.cn}
\affiliation{School of Physics Science and Engineering, Tongji University,
Shanghai 200092, China}
 
\preprint{WUCG-26-10} 
 
\date{\today}

\begin{abstract}
We constrain a scalar-vector-tensor (SVT) dark-energy model in which the
vector sector drives the dark-energy equation of state below $-1$ at earlier
times, while a canonical scalar field subsequently drives it across $-1$
toward a present-day value above $-1$.  We confront the model with DES Year
5 supernovae, DESI DR2 baryon acoustic oscillations, and compressed cosmic
microwave background information, and then include Gold-2017
redshift-space-distortion (RSD) measurements to probe structure growth.
The background posterior assigns substantial weight to crossing histories and
gives a better best fit than the flat $\Lambda$-cold-dark-matter ($\Lambda$CDM)
model.  The RSD analysis
preserves this qualitative background evolution, constrains the clustering
amplitude, and increases the posterior support for the directional crossing
criterion.  Since the growth analysis also imposes additional perturbative
support conditions, this increase cannot be attributed to the RSD likelihood
alone.  Matched PolyChord calculations give Bayes factors larger than unity for
SVT relative to $\Lambda$CDM under both prior-volume conventions considered
here.  The quantitative Bayes factor depends on the prior normalization assigned to the regular
radiation-era branch and is larger in the RSD analysis under either
convention.  Overall, current expansion and structure-growth data are consistent 
with a stable dynamical crossing of the phantom divide in the SVT model.
\end{abstract}

%\pacs{04.50.Kd,95.30.Sf,98.80.-k}

\maketitle

%%%%%%%%%%%%%%%%%%%%%%%%%%%%%
\section{Introduction}
\label{Intro}
%%%%%%%%%%%%%%%%%%%%%%%%%%%%%

The discovery of the accelerated expansion of the Universe with Type~Ia
supernovae (SN~Ia) established dark energy (DE) as a central problem in
cosmology
\cite{SupernovaSearchTeam:1998fmf,SupernovaCosmologyProject:1998vns}.
Subsequent measurements of cosmic microwave background (CMB) anisotropies
and baryon acoustic oscillations (BAO) provided further evidence for a
late-time component with negative pressure
\cite{WMAP:2003elm,SDSS:2005xqv}. The spatially flat
$\Lambda$-cold-dark-matter ($\Lambda$CDM) model, composed of a cosmological
constant $\Lambda$ and cold dark matter (CDM), together with the standard
baryonic and radiation sectors, provides the simplest successful description
of these observations. In this framework, the DE equation of state is $w_{\rm DE}=-1$. Despite its remarkable observational success, the $\Lambda$CDM model leaves unresolved the origin of the cosmological constant and its extremely small energy scale, as well as the coincidence that DE becomes dynamically important only at relatively recent times. These theoretical
questions provide strong motivation for exploring dynamical DE models and
modifications of gravity
\cite{Peebles:2002gy,Copeland:2006wr,SilvestriTrodden:2009,CliftonEtAl:2012,Tsujikawa:2013Quintessence,JoyceEtAl:2015,Koyama:2016,BullEtAl:2016,Heisenberg:2019Review,KaseTsujikawa:2019Review}.

Recent BAO measurements from the Dark Energy Spectroscopic Instrument (DESI) 
have strengthened the observational motivation for testing a time-dependent DE 
equation of state \cite{DESI:2024mwx,DESI:2024aqx,DESI:2025zgx}. 
A commonly used phenomenological description is the Chevallier--Polarski--Linder (CPL) parametrization, $w_{\rm DE}(z)=w_0+w_a z/(1+z)$
\cite{Chevallier:2000qy,Linder:2002et}, where $z$ denotes the redshift, 
$w_0=w_{\rm DE}(0)$, and $w_a$ controls the redshift dependence of 
the DE equation of state. The DESI DR2 analysis
combined with Planck CMB information reports a $3.1\sigma$ preference
for $w_a\ne0$ over $\Lambda$CDM. After SN~Ia data are included, the
preference remains at least $2.8\sigma$, with its precise significance
depending on the supernova compilation
\cite{DESI:2025zgx,DESI:2025fii}. Of particular interest is the
preferred evolution from $w_{\rm DE}<-1$ at earlier times to
$w_{\rm DE}>-1$ at low redshift. Such a crossing of the phantom divide 
provides a concrete theoretical target, since it is important to determine 
whether this behavior can arise dynamically in a consistent theory rather 
than being imposed phenomenologically.

Realizing such an evolution within a theoretically consistent framework is nontrivial. 
A minimally coupled canonical scalar field with a standard-sign kinetic term satisfies 
$w_{\rm DE}\geq -1$ and therefore cannot cross the phantom divide
\cite{Fujii:1982,Ratra:1987rm,Wetterich:1987fm,Chiba:1997,
FerreiraJoyce:1997,CaldwellDaveSteinhardt:1998,CopelandLiddleWands:1998}.
A stable crossing of the phantom divide is also generically prohibited in single-field 
k-essence in the absence of ghost and gradient instabilities
\cite{Chiba:1999ka,Armendariz-Picon:2000nqq,Armendariz-Picon:2000ulo}.
A scalar field with a negative-sign kinetic term can instead realize $w_{\rm DE}<-1$
\cite{Caldwell:1999ew}, and combining canonical and phantom fields 
allows the equation of state to cross $w_{\rm DE}=-1$
\cite{Feng:2004ad,Guo:2004fq}. Such constructions, however, contain a ghost degree of freedom and suffer from vacuum instabilities
\cite{Carroll:2003st,Cline:2003gs}. This motivates stable mechanisms in which the phantom-divide crossing emerges dynamically without introducing a fundamental phantom field.

Scalar Galileon theories and their covariant extensions, the latter forming a subclass of Horndeski theories \cite{Horndeski:1974wa}, can sustain a ghost-free phantom-like regime with $w_{\rm DE}<-1$ \cite{Nicolis:2008in,Deffayet:2009wt,Silva:2009km,Kobayashi:2010cm,DeFelice:2010pv,DeFelice:2010nf,Deffayet:2010qz,Nesseris:2010pc}. A similar behavior arises in generalized Proca (GP) theories, a class of second-order vector-tensor theories in which the $U(1)$ gauge symmetry is broken \cite{Heisenberg:2014rta,Tasinato:2014eka,Allys:2015sht,BeltranJimenez:2016rff,Allys:2016jaq}. Their cosmological solutions can evolve in the regime $w_{\rm DE}<-1$ without ghost instabilities and asymptotically approach a de Sitter state \cite{DeFelice:2016yws,DeFelice:2016uil,deFelice:2017paw}. In shift-symmetric theories with luminal tensor propagation, however, a stable crossing of $w_{\rm DE}=-1$ is obstructed by a no-go result \cite{Tsujikawa:2025wca,Aoki:2021wew,Aoki:2024ktc,Aoki:2025bmj}.

This obstruction can be evaded in scalar-tensor (ST) theories by breaking the shift symmetry through a nonconstant scalar-field potential \cite{Tsujikawa:2025wca}. A representative example is the Galileon ghost-condensate model, for which a stable phantom-divide crossing is forbidden in the shift-symmetric case \cite{Peirone:2019aua}. Once a potential is introduced, stable crossing solutions can arise, while both ghost and Laplacian instabilities are avoided throughout the cosmological evolution from the radiation-dominated era to the future accelerating attractor in the viable parameter space \cite{Tsujikawa:2025wca}.

In the vector-tensor sector, an analogous mechanism can be realized by extending GP theories through the inclusion of a canonical scalar field with a potential. The model proposed in Ref.~\cite{Tsujikawa:2026xqm}, which is the subject of the present study, provides such a realization within the scalar-vector-tensor (SVT) class \cite{Heisenberg:2018acv,Heisenberg:2018mxx,Kase:2018nwt,Heisenberg:2018wye}. In this model, the vector field leads to a DE equation of state in the regime $w_{\rm DE}<-1$ at earlier times, while the canonical scalar field, as it rolls along its potential, subsequently induces a crossing of the phantom divide toward $w_{\rm DE}>-1$.

Unlike recent scenarios invoking nonminimal couplings to account for the DESI phantom-crossing phenomenology \cite{Ye:2024desi,Wolf:2024nmc,Ye:2024tg,Pan:2025desi,Wolf:2025nmc,Wang:2025nmc,Adam:2025nmc,SanchezLopez:2025nmc}, the present SVT model contains no nonminimal coupling of either the scalar or vector field to the Ricci scalar or Einstein tensor. It thus avoids the fifth-force effects and direct time variation of the effective gravitational coupling associated with such curvature couplings, which are tightly constrained by local gravity experiments \cite{Babichev:2011iz,Kimura:2011dc,Hofmann:2018myc,Tsujikawa:2019pih}. In addition, the vector derivative interactions are restricted to cubic order, ensuring a luminal tensor propagation speed. The theoretical analysis of Ref.~\cite{Tsujikawa:2026xqm} demonstrated the existence of parameter regions free of both ghost and Laplacian instabilities throughout the cosmological evolution, from the radiation-dominated era to the future accelerating attractor.

The early-time behavior of the ST model of Ref.~\cite{Tsujikawa:2025wca} is qualitatively different. As shown in Ref.~\cite{Pookkillath:2026kinetic}, when the solution is evolved back to high redshift, the DE equation of state becomes positive, approaching $w_{\rm DE}\simeq 1/6$ during radiation domination, while the cubic Galileon contribution develops a transient braiding peak around radiation--matter equality. Although the DE density remains subdominant, this transient contribution can have a non-negligible impact on cosmological perturbations. 
The associated Galileon braiding modifies the pre-recombination gravitational driving and leaves observable imprints on the CMB anisotropy spectra \cite{Pookkillath:2026kinetic}, which have been measured with high precision in both temperature and polarization by Planck \cite{Planck:2019nip}.

By contrast, in the SVT model of Ref.~\cite{Tsujikawa:2026xqm}, the vector-field density becomes negligible toward high redshift, while Hubble friction suppresses the scalar kinetic energy and effectively freezes the scalar field on its potential, leading to $w_{\rm DE}\to -1$ in the early Universe. The background evolution therefore approaches that of $\Lambda$CDM at high redshift, while the early-time perturbation dynamics are also expected to remain close to the $\Lambda$CDM behavior, without the Galileon-braiding episode around radiation--matter equality that characterizes the ST model. The ST and SVT constructions thus exhibit qualitatively different early-time evolution at both the background and perturbation levels. In particular, the absence of a significant early-time braiding effect in the SVT model provides a well-motivated basis for employing a compressed CMB treatment based on distance priors (shift parameters) or closely related early-Universe compressed quantities \cite{Wang:2007shift,Chen:2018distance,Lemos:2023xhs} in the present statistical analysis, instead of the full Planck CMB power-spectrum likelihood \cite{Planck:2019nip}. A full Boltzmann CMB analysis nevertheless remains an important future test.

In this work, we perform a statistical analysis of the SVT model of Ref.~\cite{Tsujikawa:2026xqm} based on current cosmological observations. We first constrain the background evolution using geometric probes of the cosmic expansion. BAO measurements constrain the expansion rate and transverse comoving distance relative to the sound horizon, while SN~Ia determine the luminosity-distance--redshift relation. We further include CMB shift parameters, which provide compressed information on the distance to last scattering and the acoustic scale. Our baseline likelihood therefore combines DESI DR2 BAO \cite{DESI:2025zgx}, DES Year~5 SN~Ia \cite{DES:2024jxu}, and the compressed CMB likelihood based on Ref.~\cite{Lemos:2023xhs}. These data allow us to constrain the background parameter space of the SVT model and reconstruct the evolution of $w_{\rm DE}(z)$, thereby testing whether a stable crossing of the phantom divide is consistent with the observed expansion history.

The growth of cosmic structure provides an important complementary test of the SVT model. On scales well inside the Hubble and sound horizons, the evolution of matter perturbations is governed by an effective gravitational coupling that generally differs from its general-relativistic value. The quantities $\mu$ and $\Sigma$, which characterize matter clustering and light deflection, respectively \cite{Amendola:2007rr,Zhao:2010mg,Song:2011mg,Simpson:2012ra}, depend on the propagation properties of the longitudinal vector mode in the SVT model \cite{Kase:2018nwt,Tsujikawa:2026xqm}. 
In contrast to scalar Galileon theories \cite{DeFelice:2010as}, the transverse vector sector of the SVT model provides additional freedom that can keep these quantities close to their general-relativistic values, $\mu=\Sigma=1$ \cite{Tsujikawa:2026xqm}. The observational viability of this behavior, however, needs to be tested against measurements of structure growth. Previous cosmological studies have placed observational constraints on GP DE
models \cite{deFelice:2017paw,Nakamura:2018oyy,DeFelice:2020sdq,Heisenberg:2020xak}, but they did not incorporate the scalar-potential mechanism that gives rise to the stable phantom-divide crossing considered here. To our knowledge, a joint parameter inference for the present SVT model using both background and growth observations has not yet been performed.

Motivated by these growth-sector effects, we then extend the baseline analysis by including redshift-space distortion (RSD) measurements. RSDs arise from the line-of-sight peculiar velocities of galaxies, which induce anisotropies in the observed clustering pattern in redshift space \cite{Kaiser:1987qv}. 
On linear scales, these distortions probe the growth of cosmic structure through the combination $f\sigma_8$ \cite{Song:2008qt,Percival:2008sh}, where $f={\rm d}\ln\delta_m/{\rm d}\ln a$ is the linear growth rate of the matter density contrast $\delta_m$, $a$ is the scale factor, and $\sigma_8$ is the rms amplitude of matter fluctuations within spheres of radius $8h^{-1}\,{\rm Mpc}$. The dimensionless parameter $h$ is defined through $H_0=100h\,{\rm km}\,{\rm s}^{-1}\,{\rm Mpc}^{-1}$, where $H_0$ denotes the present-day Hubble constant. 
We use the 18-point Gold-2017 RSD compilation \cite{Nesseris:2017vor} in our growth analysis.
This enables us to assess whether the SVT parameter region that fits the background observations and realizes a stable phantom-divide crossing is also consistent with measurements of large-scale structure growth.

The background data favor a present-day equation of state above $-1$ and allow histories that cross the phantom divide at redshifts of order unity. Adding RSD data provides an independent test of the perturbation sector and further constrains the growth histories, while preserving consistency with the phantom-divide crossing solutions favored by the background probes.
The SVT model can provide a better fit than $\Lambda$CDM. We quantify this improvement using complementary model-comparison criteria, including Bayesian evidence for the adopted priors. We therefore present parameter constraints and model-comparison diagnostics separately.

This paper is organized as follows. In Sec.~\ref{backsec}, we review the SVT model and its background and linear-perturbation dynamics, including the evolution of matter perturbations. Section~\ref{methodsec} describes the parameterization, numerical shooting procedure, early-time calibration, observational likelihoods, numerical growth implementation, and sampling methodology. In Sec.~\ref{constraintsec}, we present the cosmological constraints first without and then with RSD data, reconstruct the evolution of $w_{\rm DE}(z)$ and compare the SVT model with flat $\Lambda$CDM. Finally, Sec.~\ref{consec} summarizes the results. 
Throughout this paper, we use natural units with $c=\hbar=1$, except in the definitions of cosmological distances, where the speed of light $c$ is explicitly restored.

%%%%%%%%%%%%%%%%%%%%%%%%%%%%%%%%%%%%%%%
\section{Background and linear perturbations}
\label{backsec}
%%%%%%%%%%%%%%%%%%%%%%%%%%%%%%%%%%%%%%%%%%%%%%

The SVT theories considered in this paper involve
a vector field $A_\mu$ and a scalar field $\phi$.
The vector sector contains the standard Maxwell term
$F=-(1/4)F_{\mu\nu}F^{\mu\nu}$, where
$F_{\mu\nu}=\partial_\mu A_\nu-\partial_\nu A_\mu$,
together with the terms $G_2(X)$ and
$G_3(X)\nabla_\mu A^\mu$.
Here, $G_2$ and $G_3$ are functions of
$X=-(1/2)A_\mu A^\mu$, and $\nabla_\mu$ denotes the covariant derivative.
The scalar field has the canonical kinetic term
$-(1/2)\nabla^\mu\phi\nabla_\mu\phi$ and a potential $V(\phi)$.
The gravitational sector is described by the Einstein-Hilbert term
$(\Mpl^2/2)R$, where $\Mpl$ is the reduced Planck mass
and $R$ is the Ricci scalar.
The resulting action is given by \cite{Tsujikawa:2026xqm}
\begin{align}
\mathcal{S} &= \int {\rm d}^4 x \sqrt{-g} \biggl[
\frac{\Mpl^2}{2} R + F + G_2(X) + G_3(X)\nabla_\mu A^\mu \nonumber \\
&\qquad\qquad\qquad~
-\frac{1}{2}\nabla^\mu\phi\nabla_\mu\phi - V(\phi)
\biggr] + \mathcal{S}_M\,,
\label{action}
\end{align}
where $g$ is the determinant of the metric tensor $g_{\mu\nu}$ and
$\mathcal{S}_M$ denotes the matter action. We assume that matter is minimally
coupled to the metric $g_{\mu\nu}$.

In the absence of the scalar field, the action~(\ref{action}) reduces to
a subclass of GP theories
\cite{Heisenberg:2014rta,Tasinato:2014eka,Allys:2015sht,BeltranJimenez:2016rff,Allys:2016jaq},
in which the $U(1)$ gauge symmetry is explicitly broken by the $X$
dependence of $G_2$ and $G_3$. On a spatially flat
Friedmann-Lema\^{\i}tre-Robertson-Walker (FLRW)
background, GP theories admit solutions in which the DE equation of state,
$w_{\rm DE}$, can satisfy $w_{\rm DE}<-1$ before approaching the de Sitter
attractor with $w_{\rm DE}=-1$
\cite{DeFelice:2016yws,DeFelice:2016uil,deFelice:2017paw}.
A stable crossing of the phantom divide is nevertheless obstructed in the
corresponding shift-symmetric class
\cite{Tsujikawa:2025wca,Aoki:2021wew,Aoki:2024ktc,Aoki:2025bmj}.
In the SVT construction studied here, the addition of a canonical scalar
field with a nonconstant potential breaks the scalar shift symmetry and
provides the late-time dynamics needed to cross $w_{\rm DE}=-1$
\cite{Tsujikawa:2025wca}.

In this paper, we consider the exponential potential
\begin{equation}
V(\phi)=V_0 e^{-\lambda\phi/\Mpl}\,,
\end{equation}
where $V_0$ and $\lambda$ are constants.
As the scalar field begins to evolve along this potential at low
redshifts, the DE equation of state can cross the phantom divide
$w_{\rm DE}=-1$.

\subsection{Background cosmology}
\label{backgroundtheory}

To be concrete, let us revisit the cosmological dynamics on the
spatially flat FLRW background described by the line element
\be
{\rm d}s^2=-{\rm d}t^2+a^2(t)\delta_{ij}
{\rm d}x^i{\rm d}x^j\,,
\label{backmet}
\ee
where $a(t)$ is the scale factor and $t$ is cosmic time.
The vector field has only a temporal component $\chi(t)$,
\be
A^\mu=\left[\chi(t),0,0,0\right]\,,
\ee
while the scalar field is homogeneous, $\phi=\phi(t)$.
Varying the action~(\ref{action}) with respect to $\chi$ and $\phi$
gives
\ba
& &
\chi\left(G_{2,X}+3H\chi G_{3,X}\right)=0\,,
\label{back1}\\
& &
\ddot{\phi}+3H\dot{\phi}+V_{,\phi}=0\,,
\label{back2}
\ea
where $H\equiv\dot{a}/a$ is the Hubble expansion rate and a dot
denotes a derivative with respect to $t$. We use the notation
$G_{i,X}\equiv {\rm d}G_i/{\rm d}X$ for $i=2,3$ and
$V_{,\phi}\equiv {\rm d}V/{\rm d}\phi$.
Equation~(\ref{back1}) admits two branches:
(i) $\chi=0$ and
(ii) $G_{2,X}+3H\chi G_{3,X}=0$.
We focus on branch (ii), which allows $w_{\rm DE}<-1$ to be realized
during an intermediate stage of cosmic evolution before the late-time scalar
dynamics becomes important. Branch (i), by contrast,
reduces to quintessence at the background level, since the vector field
does not contribute to the background dynamics.

For the matter action $\mathcal{S}_M$ in Eq.~(\ref{action}), we consider
nonrelativistic matter with energy density $\rho_m$ and negligible pressure,
together with radiation with energy density $\rho_r$ and pressure
$P_r=\rho_r/3$. Varying the action~(\ref{action}) with respect to
$g_{\mu\nu}$ gives the gravitational field equations
\ba
3\Mpl^2 H^2 &=& \rho_{\rm DE}+\rho_m+\rho_r\,,
\label{back3}\\
-2\Mpl^2\dot{H}
&=& \rho_{\rm DE}+P_{\rm DE}+\rho_m
+\frac{4}{3}\rho_r\,,
\label{back4}
\ea
where the effective DE energy density and pressure are
\ba
\rho_{\rm DE} &=& -G_2+\frac{1}{2}\dot{\phi}^2+V(\phi)\,,\\
P_{\rm DE} &=& G_2-G_{3,X}\dot{\chi}\chi^2
+\frac{1}{2}\dot{\phi}^2-V(\phi)\,.
\ea
The DE equation of state is then defined by
\be
w_{\rm DE}\equiv\frac{P_{\rm DE}}{\rho_{\rm DE}}
=-1+\frac{G_{3,X}\dot{\chi}\chi^2-\dot{\phi}^2}
{G_2-\dot{\phi}^2/2-V}\,.
\label{wDE0}
\ee

For the vector-field couplings $G_2(X)$ and $G_3(X)$, we adopt
the power-law forms
\be
G_{2}(X)=b_{2}X^{p_{2}}\,,\qquad
G_{3}(X)=b_{3}X^{p_{3}}\,,
\label{G23}
\ee
where $b_2$, $b_3$, $p_2$, and $p_3$ are constants.
On the branch $G_{2,X}+3H\chi G_{3,X}=0$ in
Eq.~(\ref{back1}), these functions lead to
\be
\chi^p H=-\frac{2^{p_{3}-p_{2}}b_{2}p_2}
{3b_{3}p_{3}}\,,
\label{chipre}
\ee
where
\be
p\equiv 1-2p_2+2p_3\,.
\label{pdef}
\ee
Provided that $b_2p_2\neq 0$ and $b_3p_3\neq 0$, the
right-hand side of Eq.~(\ref{chipre}) is a finite and nonvanishing 
constant.
Without loss of generality, we consider the branch for which
$\chi^p H$ is positive and $\chi>0$ throughout the cosmological
evolution. For the vector field to contribute as a source of DE,
we require
\be
p>0\,,
\label{pcon}
\ee
so that $\chi\propto H^{-1/p}$ grows as $H$ decreases.

The vector-field energy density is
$\rho_{\chi}=-b_2(\chi^2/2)^{p_2}$, whose positivity requires
$b_2<0$. Introducing a positive constant $m$, we write
\be
b_2=-m^2\Mpl^{2(1-p_2)}\,,
\ee
so that
$\rho_{\chi}=m^2\Mpl^2[\chi^2/(2\Mpl^2)]^{p_2}$.
The corresponding density parameter is
\be
\Omega_{\chi}\equiv\frac{\rho_{\chi}}{3\Mpl^2H^2}
=\frac{2^{-p_2}m^2u^{2p_2}}{3H^2}\,,
\label{Omechi}
\ee
where $u\equiv\chi/\Mpl$.

For the scalar field, we introduce the dimensionless variables
\be
x\equiv\frac{\dot{\phi}}{\sqrt{6}\Mpl H}\,,\qquad
y\equiv\frac{\sqrt{V}}{\sqrt{3}\Mpl H}\,,
\label{xy}
\ee
with the corresponding density parameter
$\Omega_{\phi}=x^2+y^2$.
We also define the density parameters of nonrelativistic matter
and radiation as
$\Omega_m\equiv\rho_m/(3\Mpl^2H^2)$ and
$\Omega_r\equiv\rho_r/(3\Mpl^2H^2)$, respectively.
The Hamiltonian constraint~(\ref{back3}) then gives
\be
\Omega_m=1-\Omega_{\rm DE}-\Omega_r\,,
\label{Omem}
\ee
where the DE density parameter is
\be
\Omega_{\rm DE}=\Omega_{\chi}+\Omega_{\phi}
=\Omega_{\chi}+x^2+y^2\,.
\ee
Introducing the parameter
\be
s\equiv\frac{p_2}{p}\,,
\ee
the term entering $P_{\rm DE}$ can be expressed as
$G_{3,X}\dot{\chi}\chi^2=-2\Mpl^2s\dot{H}\Omega_{\chi}$.
The DE equation of state~(\ref{wDE0}) then becomes
\be
w_{\rm DE}=-1+
\frac{2(s\epsilon_H\Omega_{\chi}+3x^2)}
{3(\Omega_{\chi}+x^2+y^2)}\,,
\label{wde}
\ee
where
\be
\epsilon_H\equiv\frac{\dot{H}}{H^2}
=-\frac{3+3x^2-3y^2-3\Omega_{\chi}+\Omega_r}
{2(1+s\Omega_{\chi})}\,.
\label{epsilonH}
\ee
Here, Eq.~(\ref{back4}) has been used to derive Eq.~(\ref{epsilonH}).
The effective equation of state governing the background evolution is
\be
w_{\rm eff}=-1-\frac{2}{3}\epsilon_H\,.
\ee
Hence, the condition for cosmic acceleration,
$w_{\rm eff}<-1/3$, is equivalent to $\epsilon_H>-1$.

Since the vector and scalar fields obey Eqs.~(\ref{chipre}) and
(\ref{back2}), respectively, their background evolution 
can be described by the autonomous system
\ba
\Omega_{\chi}' &=& -2(s+1)\epsilon_H\Omega_{\chi}\,,
\label{auto1}\\
x' &=& -3x+\frac{\sqrt{6}}{2}\lambda y^2-\epsilon_H x\,,
\label{auto2}\\
y' &=& -\frac{\sqrt{6}}{2}\lambda xy-\epsilon_H y\,,
\label{auto3}\\
\Omega_r' &=& -2(2+\epsilon_H)\Omega_r\,,
\label{auto4}
\ea
where a prime denotes differentiation with respect to
${\cal N}=\ln a$.

The system admits the following cosmologically relevant fixed points:
\begin{itemize}
\item
(A) $(\Omega_{\chi},x,y,\Omega_r)=(0,0,0,1)$,~~
$w_{\rm eff}=1/3$,
\item
(B) $(\Omega_{\chi},x,y,\Omega_r)=(0,0,0,0)$,~~
$w_{\rm eff}=0$,
\item
(C) $(\Omega_{\chi},x,y,\Omega_r)=(1,0,0,0)$,~~
$w_{\rm eff}=-1$.
\end{itemize}
These fixed points correspond to the radiation-, matter-, and
DE-dominated epochs, respectively. A viable cosmological trajectory
follows the sequence
(A) $\to$ (B) $\to$ (C).
We henceforth focus on $s>0$, which is also the ghost-free condition
for the longitudinal vector mode discussed in
Sec.~\ref{perturbationtheory}. During an epoch in which $\epsilon_H$
is approximately constant
(e.g., $\epsilon_H=-2$ in the radiation era,
$\epsilon_H=-3/2$ in the matter era, and $\epsilon_H=0$ in the
de Sitter era), Eq.~(\ref{auto1}) gives
$\Omega_{\chi}\propto a^{-2(s+1)\epsilon_H}$ and hence
$\rho_{\chi}\propto a^{-2s\epsilon_H}$.
For $\epsilon_H<0$, as is the case during the radiation- and matter-dominated eras, $\rho_{\chi}$ therefore increases with time for $s>0$.
In the limit $x\to0$ and $y\to0$, Eq.~(\ref{wde}) reduces to
$w_{\rm DE}=-1+2s\epsilon_H/3$. Hence, when the vector field dominates
the DE sector, the conditions $s>0$ and $\epsilon_H<0$ lead to
$w_{\rm DE}<-1$.

Since we are interested in the case where the scalar-field potential
contributes to the late-time dynamics and induces the phantom-divide
crossing, its characteristic energy scale is
$V\sim \Mpl^2H_0^2$, where $H_0$ is the present-day value of $H$.
For the exponential potential, the scalar mass squared,
$m_\phi^2\equiv V_{,\phi\phi}$, is therefore of order
$m_\phi^2\sim\lambda^2H_0^2$. Hence, for $\lambda$ at most of order unity,
$m_\phi^2\ll H^2$ during the radiation and matter eras.
The scalar field is then nearly frozen, so that $V$ is approximately
constant and the regular solution for its velocity behaves as
$\dot{\phi}\propto H^{-1}$. Consequently, during the radiation and
matter eras,
$x^2\propto H^{-4}\propto t^4$ and
$y^2\propto H^{-2}\propto t^2$, rather than $y^2$ approaching a
constant. The corresponding regular radiation-era relation,
$x=(\sqrt{6}/10)\lambda y^2$, is derived in
Sec.~\ref{shootingmethod}. Since $x^2/y^2\to0$ and
$\Omega_\chi/y^2\propto H^{-2s}\to0$ in the asymptotic past for $s>0$,
Eq.~(\ref{wde}) gives $w_{\rm DE}\to-1$. The side from which
$w_{\rm DE}$ approaches $-1$ depends on the relative contributions
of the scalar and vector sectors.

For the crossing histories of interest, $w_{\rm DE}$ is therefore close
to $-1$ deep in the radiation-dominated epoch. As the vector-field
contribution grows relative to the potential energy, the evolution
enters a regime in which
\be
w_{\rm DE}\simeq -1+
\frac{2s\epsilon_H\Omega_\chi}
{3(\Omega_\chi+y^2)}<-1\,.
\ee
At low redshifts, the scalar field begins to evolve along the potential,
and its kinetic contribution $x^2$ can no longer be neglected in
Eq.~(\ref{wde}). The phantom-divide crossing occurs at
\be
3x^2=-s\epsilon_H\Omega_\chi\,.
\ee
After the crossing, $3x^2>-s\epsilon_H\Omega_\chi$, so that
$w_{\rm DE}>-1$.
We focus on parameter regions for which this transition can occur by
the present epoch, while whether such crossing histories are favored
by observations is determined by the statistical analysis rather than
imposed as a prior condition.

After crossing the phantom divide, $w_{\rm DE}$ reaches a maximum and
subsequently evolves toward the de Sitter fixed point (C), for which
$w_{\rm DE}=-1$. There also exists another fixed point,
(D)~$(\Omega_\chi,x,y,\Omega_r)
=(0,\lambda/\sqrt{6},\sqrt{1-\lambda^2/6},0)$,
which exists for $\lambda^2\leq 6$ and has
$w_{\rm eff}=-1+\lambda^2/3$. Hence, it supports cosmic acceleration
for $\lambda^2<2$. However, this point is unstable against homogeneous
perturbations \cite{Tsujikawa:2026xqm}, and the solutions ultimately
approach the stable de Sitter point (C).
Thus, the asymptotic value of $w_{\rm DE}$ is $-1$ in both the past
and future limits.

\subsection{Linear perturbations and matter growth}
\label{perturbationtheory}

We next summarize the linear perturbation equations relevant to the
growth analysis. The linear density contrast of nonrelativistic matter
is defined as $\delta_m\equiv\delta\rho_m/\rho_m$, where $\rho_m$
is the homogeneous background energy density and $\delta\rho_m$
is its linear perturbation. On scales well inside both the Hubble radius
and the sound horizon of the longitudinal vector mode, the quasi-static
approximation gives \cite{Tsujikawa:2026xqm}
\be
\delta_m''+(2+\epsilon_H)\delta_m'
-\frac{3}{2}\mu\Omega_m\delta_m=0\,,
\label{growthequation}
\ee
where the effective gravitational coupling for matter perturbations is
\be
\mu=1+\frac{Q_\chi}{3(1+Q_\chi)c_\psi^2}\,,
\qquad Q_\chi\equiv s\Omega_\chi\,.
\label{mueffective}
\ee
For the minimal vector sector considered here, the light-deflection
parameter is also given by $\Sigma=\mu$ \cite{Tsujikawa:2026xqm}.
Hence, departures from the general-relativistic values
$\mu=\Sigma=1$ are controlled by the longitudinal-vector sound speed
$c_\psi$.

Although the background evolution depends on the vector exponents
$p_2$ and $p_3$ only through $s=p_2/p$, the perturbation sector retains
an independent dependence on the exponent $p$ defined in
Eq.~(\ref{pdef}). It is also useful to introduce the constant
\be
\nu \equiv u^p \frac{H}{m}\,,
\label{nudef}
\ee
which is time independent on the branch~(\ref{chipre}), together with
the combination
\be
\nu_v\equiv q_v\nu^{2/[p(1+s)]}\,.
\label{nuvdef}
\ee
Here, $q_v$ denotes the kinetic coefficient of the transverse vector
perturbations, which is fixed to unity in the model defined by
Eq.~(\ref{action}). At fixed $s$, the parameters $p$ and $\nu_v$
provide additional freedom in the perturbation dynamics.

For the model~(\ref{G23}), the squared propagation speed of the
longitudinal vector mode can be written as
\be
c_\psi^2=\frac{{\cal N}_\psi}{6p^2(1+Q_\chi)^2}
+\frac{2s \Omega_\chi^{1-\beta}}
{3(1+Q_\chi)2^{s/(1+s)}3^\beta\nu_v},
\label{cs2psi}
\ee
where
\ba
\beta &=&\frac{1}{p(1+s)},
\label{betadef}\\
{\cal N}_\psi &=&6ps+5p-3
+[3-3p-2ps(2+p)]\Omega_\chi-2p^2Q_\chi^2
\nonumber\\
& &+(2ps+p-1)(\Omega_r+3x^2-3y^2).
\label{npsidef}
\ea
For the longitudinal vector mode, we impose the conditions
\cite{Tsujikawa:2026xqm}
\be
s>0,\qquad ps\leq1,\qquad p(1+s)\geq1\,.
\label{rsdtheorysupport}
\ee
The first condition ensures the absence of ghosts, while the remaining
inequalities avoid early-time strong-coupling limits and keep the
sound-speed expression regular. 
We additionally require the squared sound speed to remain positive,
\be
c_\psi^2({\cal N})>0
\quad\hbox{along the cosmological history},
\label{positivecs2}
\ee
thereby excluding Laplacian instabilities.

For $\Omega_\chi>0$, these conditions imply $Q_\chi=s\Omega_\chi>0$,
so that Eq.~(\ref{mueffective}) gives $\mu>1$.
The deviation from the general-relativistic value is controlled by both $Q_\chi$
and $c_\psi^2$, and $\mu$ remains close to unity when
$Q_\chi/[3(1+Q_\chi)c_\psi^2]\ll1$.
These perturbative conditions are imposed in the RSD analysis
described in Sec.~\ref{rsdmethod}.

%%%%%%%%%%%%%%%%%%%%%%%%%%%%%%%%%%%%%%
%%%%%%%%%%%%%%%%%%%%%%%%%%%%%%%%%%%%%%
\section{Numerical and statistical methodology}
\label{methodsec}
%%%%%%%%%%%%%%%%%%%%%%%%%%%%%%%%%%%%%%

This section describes the numerical and statistical framework used in the
two observational analyses presented below. Section~\ref{backgroundresults}
presents the background constraints obtained from 
DES Year 5 SN~Ia, DESI DR2 BAO, and the three-parameter compressed CMB likelihood, while
Sec.~\ref{rsdresults} extends this baseline 
by including the RSD
measurements of $f\sigma_8$. We first describe the common parameterization,
background solver, shooting procedure, early-time calibration, and distance
likelihoods, and then specify the numerical growth integration and RSD
likelihood based on the perturbation equations of
Sec.~\ref{perturbationtheory}.

Both analyses are performed with the Metropolis MCMC sampler implemented
in \textsc{Cobaya} \cite{Torrado:2020dgo,Lewis:2002ah,Lewis:2013hha},
and the marginalized distributions are analyzed with \textsc{GetDist}
\cite{Lewis:2019xzd}. Our numerical implementation is based on a modified
version of the COSMODS \cite{Roy:2026icy} package, adapted to the background and
linear-perturbation equations of the present SVT model. 
For the baseline analysis, we use the public \texttt{sn.desy5}
likelihood based on the DES Year~5 SN~Ia sample
\cite{DES:2024jxu}, without adding Pantheon+ or a local
absolute-distance calibration. DESI DR2 is included only through
its BAO likelihood \cite{DESI:2025zgx}, without full-shape or
DESI-RSD information.
Throughout this section, a subscript 0 denotes the present
epoch, and we set $a_0=1$.

\subsection{Physical parameters and priors}
\label{physicalpriors}

It is useful to characterize the present composition of DE by introducing
the fraction $f_\chi$ of the total DE density contributed by 
the vector field,
\be
f_\chi \equiv \frac{\Omega_{\chi0}}{\Omega_{{\rm DE},0}},
\qquad
\Omega_{{\rm DE},0}=1-\Omega_{m0}-\Omega_{r0}\,.
\label{fchidef}
\ee
In terms of $f_\chi$, the present-day density parameter of the scalar
field is given by
\be
\Omega_{\phi0}=(1-f_\chi)\Omega_{{\rm DE},0}\,.
\ee

Instead of sampling $s$ directly, we introduce the combination
\be
s_\chi\equiv s f_\chi.
\label{schidef}
\ee
The background evolution depends on $s$ through combinations such as
$Q_\chi=s\Omega_\chi$. At the present epoch, this quantity is
\be
Q_{\chi0}\equiv s\Omega_{\chi0}
=s_\chi\Omega_{{\rm DE},0}.
\label{Qchi0def}
\ee
Hence, $s_\chi$ directly characterizes the impact of the vector sector on
the late-time dynamics. When $f_\chi$ is small, the background evolution
becomes weakly sensitive to $s$ itself, and a large value of $s$ need not
correspond to a large vector-field effect. Sampling $s_\chi$ therefore
avoids introducing an independent upper cutoff on $s$ that would truncate
this degeneracy according to an arbitrarily chosen exponent range.

We consider two observational analyses: a baseline analysis without RSD
data, which we refer to as the no-RSD analysis, and an extended analysis
including RSD data, referred to as the RSD analysis. The six parameters
common to both analyses are
\begin{equation}
\boldsymbol{\vartheta}_{\rm bg}
=\left(H_0,\Omega_{b0},\Omega_{m0},f_\chi,s_\chi,\lambda\right),
\label{backgroundparameters}
\end{equation}
where
$\Omega_{b0}\equiv\rho_{b0}/(3\Mpl^2H_0^2)$ and
$\Omega_{m0}\equiv\rho_{m0}/(3\Mpl^2H_0^2)$ are the present-day density
parameters of baryons and total nonrelativistic matter, respectively.
The no-RSD analysis samples only the parameter vector
$\boldsymbol{\vartheta}_{\rm bg}$.

For the RSD analysis, we enlarge the parameter space to
\begin{equation}
\boldsymbol{\vartheta}_{\rm RSD}
=\left(\boldsymbol{\vartheta}_{\rm bg},p,\xi_v,\sigma_{8,0}\right),
\qquad
\xi_v\equiv\log_{10}\nu_v,
\label{rsdparameters}
\end{equation}
where $p$ is the vector power-law parameter defined in
Eq.~(\ref{pdef}), $\nu_v$ is the perturbation-sector combination
defined in Eq.~(\ref{nuvdef}), and $\sigma_{8,0}$ denotes the present-day
rms amplitude of matter fluctuations on the scale
$8h^{-1}\,{\rm Mpc}$. The quantities
$s=s_\chi/f_\chi$ and $\nu_v=10^{\xi_v}$ are derived at every sampled point.

The physical priors are given in
Table~\ref{tab:svt_priors} and are defined in terms of the physical
variables listed there. The no-RSD chains sample the six background
quantities in Eq.~(\ref{backgroundparameters}) directly. For the RSD
analysis, an equivalent reparameterization is used for numerical efficiency,
with the corresponding Jacobian included so that the induced prior in the
physical variables remains that of Table~\ref{tab:svt_priors}.
The formal priors for $f_\chi$ and $s_\chi$ span $[0,1]$. The background
solver evaluates the regular domain $0<f_\chi<1$ and
$0\leq s_\chi<1$. The excluded boundary points have zero measure under the
continuous priors.

%%%%%%%%%%%%%%%%%%%%%%%%%%%%
\begin{table*}[t]
\centering
\caption{Sampled parameters and priors. The six background quantities are
common to the analyses without and with RSD. The final three quantities
are activated only for the RSD analysis. 
The $H_0$ prior is expressed in ${\rm km\,s^{-1}\,Mpc^{-1}}$, while all
other sampled quantities are dimensionless. Background regularity and the
early-Universe requirements in Eqs.~(\ref{calibrationdomain}) and
(\ref{earlyDEcut}) are imposed in both the RSD and no-RSD analyses. The RSD
analysis is further restricted by the perturbative support conditions in
Eqs.~(\ref{rsdtheorysupport}) and (\ref{positivecs2}).
}
 \label{tab:svt_priors}
 \begin{ruledtabular}
 \begin{tabular}{llll}
 Parameter & Definition & Physical prior & Usage \\
 \hline
 $H_0$ & Present Hubble expansion rate & ${\cal U}(60,80)$ & Both \\
 $\Omega_{b0}$ & Present baryon density fraction & ${\cal U}(0.03,0.07)$ & Both \\
 $\Omega_{m0}$ & Present baryon plus CDM fraction
 & ${\cal U}(0.20,0.45)$ & Both \\
 $f_\chi$ & $\Omega_{\chi0}/\Omega_{{\rm DE},0}$
 & ${\cal U}(0,1)$ & Both \\
 $s_\chi$ & $s f_\chi$ & ${\cal U}(0,1)$ & Both \\
 $\lambda$ & Exponential-potential slope & ${\cal U}(10^{-3},10)$ & Both \\
 $p$ & Vector power-law parameter in Eq.~(\ref{pdef}) & ${\cal U}(0,25)$ & RSD only \\
 $\xi_v$ & $\log_{10}\nu_v$ with Eq.~(\ref{nuvdef}) & ${\cal U}(-3,2)$ & RSD only \\
 $\sigma_{8,0}$ & Present fluctuation amplitude & ${\cal U}(0.5,1.1)$ & RSD only \\
 \end{tabular}
 \end{ruledtabular}
\end{table*}
%%%%%%%%%%%%%%%%%%%%%%%%%%%%%%

A flat prior in $(s_\chi,f_\chi)$ is not equivalent to a flat prior in
$(s,f_\chi)$. Indeed, the Jacobian determinant is given by 
\begin{equation}
 J\equiv
 \left|
 \frac{\partial(s_\chi,f_\chi)}
 {\partial(s,f_\chi)}
 \right|
 =f_\chi\,.
 \label{schipriormeasure}
\end{equation}
Hence, a flat prior in $(s_\chi,f_\chi)$ induces a density proportional to
$f_\chi$ in the $(s,f_\chi)$ coordinates, within the transformed support.
No independent upper prior is imposed on
$s=s_\chi/f_\chi$; in particular, we do not introduce an arbitrary upper
cutoff on $s$ that would truncate the small-$f_\chi$ region. The conditions
$s_\chi\geq0$ and $f_\chi>0$ imply $s\geq0$. We also do not impose any
prior requiring a phantom-divide crossing.

In the no-RSD analysis, the background evolution depends on the vector
exponents only through the combination $s=p_2/p$. Consequently, $p$,
$\nu_v$, and $\sigma_{8,0}$ are not sampled, and no perturbative stability
cut is imposed. The resulting background-only constraints should therefore
not be interpreted as a test of all perturbative stability conditions of
the underlying theory.

For the RSD analysis, the uniform prior in $\xi_v$ corresponds to a
log-uniform prior on $10^{-3}<\nu_v<10^2$. In addition to the rectangular
priors in Table~\ref{tab:svt_priors}, we impose the perturbative support
conditions in Eqs.~(\ref{rsdtheorysupport}) and
(\ref{positivecs2}), derived in Sec.~\ref{perturbationtheory}.
In sampled coordinates, the analytic inequalities in
Eq.~(\ref{rsdtheorysupport}) are equivalently
$p s_\chi\leq f_\chi\leq p(s_\chi+f_\chi)$.
Because these perturbative support conditions are imposed only in the RSD
analysis, a direct comparison of the RSD and no-RSD posterior distributions
does not isolate the effect of adding the RSD data alone. Such a comparison
would require the no-RSD control to be reweighted or rerun with the same
perturbative support. This distinction is relevant to posterior comparisons.
The normalization of the common support is included explicitly in the
matched RSD evidence calculation of Sec.~\ref{rsdresults}.

\subsection{Algebraic reconstruction using the normalized Hubble rate}
\label{algebraicbackground}

Direct integration of $\Omega_\chi$ is numerically delicate when it
becomes exponentially small toward the past. A finite absolute
integration error can then overwhelm its physical value and produce
spurious negative values. To avoid evolving this small quantity
independently, we introduce
\begin{equation}
E({\cal N})\equiv\frac{H({\cal N})}{H_0},\qquad \ell({\cal N})\equiv\ln E({\cal N})\,,
\label{Elogdef}
\end{equation}
where  ${\cal N}=\ln a=-\ln(1+z)$.
With $E(0)=1$, Eq.~(\ref{auto1}) has the exact first integral
$\Omega_\chi=\Omega_{\chi0}E^{-2(1+s)}$. We implement this relation
in terms of the physical parameters of 
Eq.~(\ref{backgroundparameters}) as
\begin{align}
 {\cal D}({\cal N})&\equiv 2\ell({\cal N})+\frac{2s_\chi\ell({\cal N})}{f_\chi},
 \label{decaydef}\\
 \Omega_\chi({\cal N})&=f_\chi\Omega_{{\rm DE},0}e^{-{\cal D}({\cal N})},
 \label{algebraicOchi}\\
 Q_\chi({\cal N})&=s_\chi\Omega_{{\rm DE},0}e^{-{\cal D}({\cal N})}.
 \label{algebraicQchi}
\end{align}
Here $Q_\chi$ is used in place of the combination $s\Omega_\chi$ throughout, with $s$ treated as a constant model parameter and $Q_\chi=s\Omega_\chi$ determined algebraically.
Separate conservation of pressureless matter and radiation gives
\begin{equation}
\Omega_m({\cal N})=\Omega_{m0}e^{-3{\cal N}-2\ell},\qquad
\Omega_r({\cal N})=\Omega_{r0}e^{-4{\cal N}-2\ell}.
\label{algebraicfluids}
\end{equation}
For the numerical background evolution, we retain the variables $x$ and $y$
introduced in Sec.~\ref{backgroundtheory}, together with
$\ell=\ln(H/H_0)$. Thus only $(x,y,\ell)$ are integrated. The evolution
equations for $x$ and $y$ are those already given in
Eqs.~(\ref{auto2}) and (\ref{auto3}), while
$\ell'=\epsilon_H$. The Hubble slope and the DE equation of state are
evaluated from Eqs.~(\ref{epsilonH}) and (\ref{wde}), with the algebraic
substitutions
$s\Omega_\chi=Q_\chi$ and
$\Omega_{\rm DE}=\Omega_\chi+x^2+y^2$.
All other quantities entering these equations are reconstructed
algebraically from Eqs.~(\ref{algebraicOchi})--(\ref{algebraicfluids}).
Using the Friedmann constraint, Eq.~(\ref{epsilonH}) implies
$\epsilon_H\leq0$ on the physical expanding branch, and hence
$E\geq1$ toward the past. The algebraic expressions in
Eqs.~(\ref{algebraicOchi}) and (\ref{algebraicQchi}) preserve the
nonnegativity of the vector contributions, while $H=H_0e^\ell$ remains
positive by construction.

The physical prior enforces $f_\chi>0$, and the endpoint $f_\chi=0$ at fixed
nonzero $s_\chi$ is not sampled. 
For fixed $z>0$ with $E(z)>1$, $Q_\chi(z)\to0$ as
$f_\chi\to0^+$, whereas $Q_{\chi0}=s_\chi\Omega_{{\rm DE},0}$ remains
finite. Thus, for sufficiently small $f_\chi$, $Q_\chi$ varies rapidly in a
narrow region close to $z=0$, which is relevant for reconstructing the
low-redshift behavior of $w_{\rm DE}(z)$.

\subsection{Radiation-era initial conditions and forward shooting}
\label{shootingmethod}

We initialize the background evolution deep in the radiation era at
$z_i=2\times10^5$. In this regime,
$\epsilon_H\simeq-2$ and $\Omega_{\rm DE}\ll1$, so
Eqs.~(\ref{auto2}) and (\ref{auto3}) imply
$y\propto e^{2{\cal N}}$ and
$x'+x\simeq(\sqrt{6}/2)\lambda y^2$. Requiring regularity toward the past
eliminates the homogeneous scalar mode and selects
\begin{equation}
x_i=\frac{\sqrt{6}}{10}\lambda y_i^2.
\label{regularinitial}
\end{equation}
We impose this regular branch as the initial condition, rather than
introducing an independent prior on the early-time scalar velocity.

We fix the physical radiation density by adopting
$T_{\rm CMB}=2.7255\,{\rm K}$, $N_{\rm eff}=3.046$, and massless
neutrinos. Using the reduced Hubble parameter $h$ defined by
$H_0=100\,h\,{\rm km\,s^{-1}\,Mpc^{-1}}$ and denoting the present photon
density parameter by $\Omega_{\gamma0}$, the physical photon density is
$\omega_\gamma\equiv\Omega_{\gamma0}h^2$, with
\begin{align}
\omega_\gamma&=2.4728\times10^{-5}
\left(\frac{T_{\rm CMB}}{2.7255\,{\rm K}}\right)^4,
\label{omegagamma}\\
\Omega_{r0}&=\frac{\omega_\gamma}{h^2}
\left(1+0.22710731766N_{\rm eff}\right).
\label{omegarphysical}
\end{align}
In particular, $\Omega_{r0}$ is not held fixed when $H_0$ is varied.
There is no separately evolved massive-neutrino matter component in
this implementation, and $\Omega_{m0}$ comprises baryons 
and CDM.

For a trial $\eta_i=\ln y_i$, the initial value $\ell_i$ is determined
by the Friedmann constraint,
\begin{align}
 1={}&x_i^2+y_i^2
 +\Omega_{m0}e^{-3{\cal N}_i-2\ell_i}
 +\Omega_{r0}e^{-4{\cal N}_i-2\ell_i}\nonumber\\
 &+f_\chi\Omega_{{\rm DE},0}
 \exp\!\left(-2\ell_i-\frac{2s_\chi\ell_i}{f_\chi}\right)\,.
 \label{initialclosure}
\end{align}
At fixed $\eta_i$, the fluid and vector terms entering the initial constraint
decrease monotonically as $\ell_i$ increases. 
We therefore determine $\ell_i$ on the expanding branch by a bracketed
one-dimensional root solve, without introducing an additional shooting
parameter. We then integrate forward to ${\cal N}=0$ and adjust $\eta_i$ so that the present
scalar DE density satisfies
$x_0^2+y_0^2=(1-f_\chi)\Omega_{{\rm DE},0}$. To implement this condition,
we define the shooting residual
\begin{equation}
{\cal R}(\eta_i)\equiv
\ln\left[\frac{x_0^2+y_0^2}
{(1-f_\chi)\Omega_{{\rm DE},0}}\right],
\label{shootingresidual}
\end{equation}
and solve ${\cal R}(\eta_i)=0$. The shooting is therefore one dimensional,
with $\eta_i$ as the only shooting variable; the initial vector amplitude and
$\ell_i$ are fixed algebraically by the prescribed present-day parameters and
the initial constraint, respectively. The root search is initialized using an
analytic estimate based on the nearly frozen scalar potential, and a nearby
previously converged solution may be used as the starting point for numerical
efficiency.

The background equations are integrated with the DOP853 method, adopting a
relative tolerance of $2\times10^{-9}$ and componentwise absolute tolerances
$(10^{-50},10^{-40},10^{-11})$ for $(x,y,\ell)$. We require the shooting residual
$|{\cal R}|\leq2\times10^{-7}$ together with a relative scalar endpoint
error below $10^{-6}$. The final solution is sampled at 6000 points in ${\cal N}$,
with a maximum integration step of $0.05$, and the maximum deviation from the
Friedmann constraint is required to remain below $3\times10^{-5}$. 
A trial is rejected if the shooting does not converge, if the numerical
solution violates the physical background conditions, or if the resulting
early-time solution fails the required calibration and early-DE conditions.

At the final endpoint, we enforce $\ell(0)=0$ exactly. This is particularly
important for very small $f_\chi$, for which the vector contribution can vary
rapidly within a narrow region close to ${\cal N}=0$ and its relative accuracy cannot
be assessed from the absolute ODE tolerance alone. The solution at ${\cal N}<0$ is
left unchanged. This endpoint adjustment is a normalization condition rather
than a resolution test of the narrow final region, a distinction that is
relevant to the low-redshift tails discussed in
Sec.~\ref{backgroundresults}.

\subsection{Observational likelihoods and early-time calibration}
\label{datalikelihoods}

The common background likelihood and the two analysis branches are
\ba
 {\cal L}_{\rm no\text{-}RSD} 
 &=&{\cal L}_0,\qquad
 {\cal L}_{\rm with\text{-}RSD}={\cal L}_0{\cal L}_{\rm RSD},
 \label{jointlikelihood}\\
 \chi_X^2&=&-2\ln{\cal L}_X+C_X\,,
 \label{jointchi2}
\ea
where
\ba
 {\cal L}_0&=&{\cal L}_{\rm SN}{\cal L}_{\rm BAO}{\cal L}_{\rm CMB},\\
 X&\in&\{{\rm no\text{-}RSD},{\rm with\text{-}RSD}\}.
\ea
The constants $C_X$ are independent of the sampled cosmological parameters.
The supplied covariance matrix is adopted for each data set, while
cross-covariances among the supernova, BAO, compressed-CMB, and historical RSD
data sets are neglected.

\subsubsection{DES Year 5 supernovae}

The DES-SN5YR likelihood \cite{DES:2024jxu} is evaluated with the public
\texttt{sn.desy5} implementation. On a spatially flat FLRW background, we
denote by $D_M(z)$ the transverse comoving distance and by $D_L(z)$ the
luminosity distance. They are given by
\begin{equation}
 D_M(z)=\frac{c}{H_0}\int_0^z\frac{\rd\tilde z}{E(\tilde z)},
 \qquad
 D_L(z)=(1+z)D_M(z).
 \label{distancedef}
\end{equation}
The corresponding theoretical distance modulus is
$\mu_{\rm th}=5\log_{10}[D_L/{\rm Mpc}]+25$, up to a common calibration
offset. The public likelihood retains the catalogue redshift conventions.
The absolute-magnitude offset is marginalized rather than calibrated by an
external distance ladder. We denote by $\chi_{\rm SN}^2$ the resulting
supernova contribution to the total chi-square, up to an additive constant
independent of the cosmological parameters. In matrix notation, let
$\boldsymbol C_{\rm SN}$ denote the supplied full supernova covariance matrix,
including statistical and systematic uncertainties, $\boldsymbol\Delta$ the
residual vector, and $\boldsymbol u\equiv(1,\ldots,1)^{\rm T}$ the
constant-offset vector. The marginalized contribution is
\begin{align}
 \chi_{\rm SN}^2
 &=\boldsymbol\Delta^{\rm T}\widetilde{\boldsymbol C}_{\rm SN}^{-1}
   \boldsymbol\Delta,\nonumber\\
 \widetilde{\boldsymbol C}_{\rm SN}^{-1}
 &=\boldsymbol C_{\rm SN}^{-1}
 -\frac{\boldsymbol C_{\rm SN}^{-1}\boldsymbol u
 \boldsymbol u^{\rm T}\boldsymbol C_{\rm SN}^{-1}}
 {\boldsymbol u^{\rm T}\boldsymbol C_{\rm SN}^{-1}\boldsymbol u}.
 \label{SNmarginalized}
\end{align}
Terms independent of the cosmological parameters do not affect the posterior.
Consequently, these supernovae primarily constrain the shape of the
distance--redshift relation rather than determining $H_0$ independently of
the other probes.

\subsubsection{DESI DR2 BAO}

We adopt the public \texttt{bao.desi\_dr2} likelihood for the DESI DR2
measurements \cite{DESI:2025zgx}. In addition to the transverse comoving
distance $D_M(z)$, we define the Hubble distance $D_H(z)$ and the
spherically averaged BAO distance $D_V(z)$ by
\begin{equation}
 D_H(z)=\frac{c}{H_0E(z)},\qquad
 D_V(z)=\left[zD_H(z)D_M^2(z)\right]^{1/3}.
 \label{BAOdistances}
\end{equation}
Here $D_H$ characterizes the radial distance scale associated with the
expansion rate, whereas $D_V$ combines the radial and transverse distance
information into the isotropic volume-averaged scale. We denote by
$r_{\rm drag}$ the comoving sound horizon at the baryon-drag epoch. The data
constrain $D_M/r_{\rm drag}$ and $D_H/r_{\rm drag}$ for anisotropic
measurements, or $D_V/r_{\rm drag}$ for isotropic measurements. The Gaussian
BAO likelihood is constructed from the measurement vector and covariance
supplied with the DESI DR2 release. Here BAO refers only to these distance
measurements and does not include an additional full-shape or RSD constraint.

\subsubsection{Compressed CMB likelihood}
\label{cmbsubsec}

CMB distance priors, often expressed in terms of shift parameters, provide a
standard compressed description of the CMB information relevant to late-time
DE constraints \cite{Wang:2007shift,Chen:2018distance}. In the present
analysis, we adopt the early-Universe compression of Appendix~A of
Ref.~\cite{DESI:2025zgx}, based on Ref.~\cite{Lemos:2023xhs}. In the order
entering our likelihood, the compressed variables are
\begin{equation}
 \boldsymbol d=(\theta_\star,\omega_b,\omega_{bc}),\qquad
 \theta_\star=\frac{r_\star}{D_M(z_\star)}.
 \label{cmbvariables}
\end{equation}
Here $z_\star$ is the redshift of photon decoupling (last scattering),
$r_\star$ is the comoving sound horizon evaluated at $z_\star$, and
$\theta_\star$ is the corresponding angular acoustic scale. The physical
density parameters are
$\omega_b=\Omega_{b0}h^2$ and
$\omega_{bc}=\Omega_{m0}h^2$. In our massless-neutrino implementation,
$\Omega_{m0}$ contains only baryons and CDM, so that $\omega_{bc}$ is the
physical baryon-plus-CDM density. This identification would have to be
modified if massive neutrinos were included in $\Omega_m$.

We evaluate
\begin{align}
 \chi^2_{\rm CMB}
 &=(\boldsymbol d-\bar{\boldsymbol d})^{\rm T}
 \boldsymbol C_{\rm CMB}^{-1}
 (\boldsymbol d-\bar{\boldsymbol d}),
 \label{cmblikelihood}\\
 \bar{\boldsymbol d}&=(0.01041,0.02223,0.14208),
 \label{cmbmean}\\
 \boldsymbol C_{\rm CMB}&=10^{-9}
 \begin{pmatrix}
  0.006621&0.12444&-1.1929\\
  0.12444&21.344&-94.001\\
  -1.1929&-94.001&1488.4
 \end{pmatrix}.
 \label{cmbcov}
\end{align}
This compression reduces sensitivity to late-time CMB lensing and the
integrated Sachs--Wolfe effect. It retains the acoustic-scale and
physical-density information relevant to the background analysis, but is not
a full SVT CMB power-spectrum likelihood and therefore does not directly test
the late-time perturbations of every background solution admitted by the
sampling chains.

\subsubsection{Normalized-hybrid sound horizons}
\label{soundcalibration}

The fitting formulae for the photon-decoupling and baryon-drag epochs and
their corresponding sound horizons are calibrated against
\textsc{CAMB} \cite{Lewis:1999bs}. For
$X\in\{\star,{\rm drag}\}$, the implementation represents the calibrated
reference values $z_X^{\rm fit}$ and $r_X^{\rm fit}$ by quadratic
functions of $(\omega_b,\omega_{bc})$. The SVT dependence of the sound
horizon is then incorporated separately through the ratio of sound-horizon
integrals evaluated with the SVT and reference expansion histories. The
quantities entering the likelihood are evaluated as
\begin{align}
 z_X&=z_X^{\rm fit}(\omega_b,\omega_{bc}),
 \label{epochfit}\\
 r_X&=r_X^{\rm fit}(\omega_b,\omega_{bc})
       \frac{I_X[H_{\rm SVT}]}{I_X[H_{\rm ref}]},
 \label{normalizedhorizon}\\
 I_X[H]&=\int_{z_X}^{\infty}\frac{c_s(z)}{H(z)}\,\rd z,
 \label{soundintegral}\\
 c_s(z)&=\frac{c}{\sqrt{3[1+R_b(z)]}},\qquad
 R_b(z)=\frac{3\omega_b}{4\omega_\gamma(1+z)}.
 \label{soundspeed}
\end{align}
Here $I_X[H]$ is the comoving sound-horizon integral from the epoch $z_X$,
$c_s$ is the sound speed of the tightly coupled photon--baryon fluid, and
$R_b$ is the baryon-to-photon momentum-density ratio. The reference
expansion history is that of a spatially flat $\Lambda$CDM model with the
same $H_0$, $\Omega_{m0}$, and $\Omega_{r0}$,
\begin{equation}
 \frac{H_{\rm ref}^2(z)}{H_0^2}
 =\Omega_{r0}(1+z)^4+\Omega_{m0}(1+z)^3
    +1-\Omega_{m0}-\Omega_{r0}.
 \label{referenceH}
\end{equation}
The same fitted value of $z_X$ and the same baryon-loading function
$R_b(z)$ enter the numerator and denominator of
Eq.~(\ref{normalizedhorizon}). Hence, when the SVT and reference expansion
histories coincide, the ratio of sound-horizon integrals is unity and the
calibrated value $r_X^{\rm fit}$ is recovered. For an SVT background, the
ratio accounts for the residual change in the expansion history. At
$z>z_i$, both integrals are continued with the same standard early-time
expansion.

The calibration is restricted to
\begin{equation}
 0.0214\leq\omega_b\leq0.0234,\qquad
 0.13\leq\omega_{bc}\leq0.15.
 \label{calibrationdomain}
\end{equation}
Points outside this domain are rejected rather than extrapolated. We further
require
\begin{equation}
 \max\left[\Omega_{\rm DE}(z_\star),
           \Omega_{\rm DE}(z_{\rm drag})\right]<10^{-4}.
 \label{earlyDEcut}
\end{equation}
These restrictions ensure that the analysis remains within the
standard-early-Universe regime for which the compressed CMB likelihood and
the calibrated sound horizons are applicable. They do not replace a full
perturbation calculation. The early-DE fractions and the two sound-horizon
integral ratios are retained as derived diagnostic quantities.

\subsection{Growth integration and the RSD likelihood}
\label{rsdmethod}

For the RSD analysis, we integrate the matter-growth equation
(\ref{growthequation}) along each accepted background history, with the
effective gravitational coupling in Eq.~(\ref{mueffective}) and the
scalar-mode propagation speed in Eq.~(\ref{cs2psi}), both derived in
Sec.~\ref{perturbationtheory}. 
Samples that violate the perturbative support conditions in
Eqs.~(\ref{rsdtheorysupport}) and (\ref{positivecs2}) are rejected before
the RSD likelihood is computed. For numerical stability, the second term in
Eq.~(\ref{cs2psi}) is handled in logarithmic form when $\Omega_\chi$
becomes very small at early times.  Since Eq.~(\ref{growthequation}) is a
quasi-static result, its application to the compiled RSD measurements assumes
that the relevant scales lie sufficiently inside the longitudinal-vector sound
horizon for the posterior region considered.  A full scale-dependent
perturbation treatment is beyond the scope of the present work.

The matter-growth equation is integrated from ${\cal N}_g=-4.5$ to the present
epoch, ${\cal N}=0$. The growing-mode initial condition is set using the
M\'esz\'aros solution for smooth radiation,
\begin{align}
 \delta_m&\propto a+\frac{2a_{\rm eq}}{3},
 \nonumber\\
 \left.\frac{\delta_m'}{\delta_m}\right|_{{\cal N}_g}
 &=\frac{a_g}{a_g+2a_{\rm eq}/3},
 \qquad
 a_g=e^{{\cal N}_g},\qquad
 a_{\rm eq}=\frac{\Omega_{r0}}{\Omega_{m0}}.
 \label{growthinitial}
\end{align}
The overall normalization of $\delta_m$ is arbitrary and cancels in the
observable below. We solve the growth equation with DOP853, using a relative
tolerance of $2\times10^{-7}$, an absolute tolerance of $10^{-10}$, and a
maximum step size of $0.03$. The redshifts of the RSD measurements are
included explicitly among the integration points. The theoretical RSD
prediction is then
\begin{equation}
 f\sigma_8(z)=\sigma_{8,0}
 \frac{\delta_m'[{\cal N}(z)]}{\delta_m(0)},
 \qquad
 {\cal N}(z)=-\ln(1+z).
 \label{fsigma8prediction}
\end{equation}

We adopt the 18-point Gold-2017 compilation over
$0.02\leq z\leq1.40$ \cite{Nesseris:2017vor}.
For measurements extracted assuming a fiducial cosmology, 
we apply the approximate geometry correction used in the compilation,
\begin{equation}
 q_i=\frac{H(z_i)D_A(z_i)}
 {H_i^{\rm fid}(z_i)D_{A,i}^{\rm fid}(z_i)},
 \qquad
 m_i=q_i[f\sigma_8]_{\rm th}(z_i),
 \label{rsdgeometry}
\end{equation}
where $H_i^{\rm fid}$ and $D_{A,i}^{\rm fid}$ are evaluated in the
fiducial cosmology employed in the original analysis of 
the $i$th measurement.
The RSD likelihood is then evaluated as
\begin{equation}
 \chi^2_{\rm RSD}
 =(\boldsymbol d_{\rm RSD}-\boldsymbol m)^{\rm T}
 \boldsymbol C_{\rm RSD}^{-1}
 (\boldsymbol d_{\rm RSD}-\boldsymbol m),
 \label{rsdlikelihood}
\end{equation}
where
$\boldsymbol d_{\rm RSD}
=\bigl([f\sigma_8]_{\rm obs}(z_1),\ldots,
[f\sigma_8]_{\rm obs}(z_{18})\bigr)^{\rm T}$
is the RSD data vector,
$\boldsymbol m=(m_1,\ldots,m_{18})^{\rm T}$ is the corresponding
theoretical prediction vector with
$m_i=q_i[f\sigma_8]_{\rm th}(z_i)$, and
$\boldsymbol C_{\rm RSD}^{-1}$ is the inverse of 
the covariance matrix
$\boldsymbol C_{\rm RSD}$ for the 18 RSD measurements.
For the three WiggleZ measurements at $z=(0.44,0.60,0.73)$, we retain
the published covariance submatrix
\begin{equation}
 \boldsymbol C_{\rm WiggleZ}=10^{-3}
 \begin{pmatrix}
 6.400&2.570&0\\
 2.570&3.969&2.540\\
 0&2.540&5.184
 \end{pmatrix},
 \label{wigglezcov}
\end{equation}
while cross-covariances between measurements from different surveys are
neglected, following the approximation adopted in the compilation. 
This likelihood uses the compiled $f\sigma_8$ measurements rather than a
reanalysis of the survey multipoles.

\subsection{Sampling strategy and convergence}
\label{samplingmethod}

For the posterior sampling, we work directly in the physical coordinates
defined above. The no-RSD analysis varies only the background parameters,
whereas the RSD analysis additionally samples $p$, $\xi_v$, and
$\sigma_{8,0}$. The target posterior density is constructed from the priors
in Table~\ref{tab:svt_priors} and the likelihood in
Eq.~(\ref{jointlikelihood}), with the perturbative support conditions in
Eqs.~(\ref{rsdtheorysupport}) and (\ref{positivecs2}) additionally imposed
in the RSD analysis. Since $\xi_v=\log_{10}\nu_v$, a uniform prior on
$\xi_v$ corresponds to a log-uniform prior on $\nu_v$. Gaussian reference
distributions centered near viable points are used solely to initialize the
sampling and do not alter the physical priors.

We distinguish between pilot chains, used to estimate the proposal
covariance, and production chains, used for the final posterior inference.
Because the sampled background parameters are strongly correlated, a
diagonal proposal covariance is inefficient. For the no-RSD production
run, we therefore use a six-dimensional proposal covariance estimated
from a pilot run with the same model, likelihood, parameterization, and
priors. For the RSD production run, a full nine-dimensional proposal
covariance is similarly estimated from an RSD pilot run in the same
physical coordinates. In both cases, the pilot runs serve only to
initialize the proposal covariance and are neither resumed nor combined
with the production posterior. The no-RSD and RSD production runs use
separate output roots, with no samples, posterior weights, burn-in
information, or checkpoints shared between them. 
In both production runs, adaptive proposal-covariance learning is enabled,
with \texttt{proposal\_scale}=1.1 setting the overall proposal scale and
\texttt{max\_tries}=2000 specifying the maximum number of proposal attempts
before a chain is considered stuck.

The posterior figures and parameter constraints presented below are obtained
from the converged MCMC production chains. PolyChord is used independently
for the matched Bayesian-evidence calculations and, where explicitly stated,
for the weighted-history crossing probability. Thus, the MCMC chains provide
the posterior samples used for marginalized constraints and reconstructed
histories, whereas PolyChord evaluates the evidence integral over the prior
volume required for model comparison. The two evidence conventions introduced
below differ only in the normalization assigned to the regular physical
support and therefore leave the normalized posterior distribution on that
support unchanged. Consequently, the MCMC contours and reconstructed
$w_{\rm DE}(z)$ histories are independent of the evidence-normalization
convention.

For each analysis, four production chains are run in parallel. \textsc{Cobaya}
discards the first 1000 accepted samples of each chain as burn-in. Unless the
trace diagnostics indicate a remaining transient, no additional fractional
burn-in is applied, and all retained samples are analyzed with their original
multiplicity weights.

Convergence is assessed using the generalized Gelman--Rubin diagnostics
implemented in \textsc{Cobaya} \cite{Lewis:2013hha}. We denote the two
reported statistics by
\begin{equation}
 \Delta_{\rm mean}\equiv (R-1)_{\rm means},
 \qquad
 \Delta_{\rm bound}\equiv (R-1)_{\rm bounds}.
 \label{GRdefinitions}
\end{equation}
The production chains are terminated when
\begin{equation}
 \Delta_{\rm mean}<0.01,
 \qquad
 \Delta_{\rm bound}<0.05.
 \label{stoppingcriteria}
\end{equation}
The first statistic monitors convergence of the multivariate means, whereas
the second is more sensitive to the stability of the credible-region
boundaries. Together, they provide a practical test of convergence, although
no finite MCMC run can exclude the existence of a disconnected mode that
remains unexplored.

\subsection{Posterior summaries and reconstruction of the equation of state}
\label{posteriormethod}

The retained samples are represented in the physical coordinates of
Eq.~(\ref{backgroundparameters}). For the RSD analysis, they additionally
include the three parameters in Eq.~(\ref{rsdparameters}) and the
corresponding derived quantities. If $W_i$ denotes the original multiplicity
weight of the $i$th retained production sample, the posterior mean of a
quantity $A$ is
\begin{equation}
 \langle A\rangle=
 \frac{\sum_i W_i A(\boldsymbol\vartheta_i)}{\sum_i W_i}.
 \label{weightedmean}
\end{equation}
The central 68\% credible interval is defined by the weighted marginalized
quantiles $[q_{0.16},q_{0.84}]$. In Table~\ref{tab:svt_constraints}, 
we report the posterior mean and the corresponding 
central 68\% credible interval. 
For a skewed posterior, the mean and central credible interval should not be
interpreted as a mode with symmetric Gaussian errors. No additional
coordinate Jacobian, weight clipping, or burn-in adjustment is applied
during post-processing.

For a curve labelled ``global best fit,'' we select the retained chain
sample with the smallest total chi-square,
\begin{align}
 \chi^2_X={}&\chi^2_{\rm SN}+\chi^2_{\rm BAO}+\chi^2_{\rm CMB}
 +I_X\chi^2_{\rm RSD},
 \label{bestfitcriterion}\\
 I_{\rm no\mbox{-}RSD}={}&0,\qquad I_{\rm RSD}=1.
 \nonumber
\end{align}
This corresponds to the best-fitting point encountered by the relevant
production chains and is not obtained from a separate continuous
minimization.

For the shaded $w_{\rm DE}(z)$ bands shown below, we draw 4000 posterior
samples from the corresponding converged MCMC posterior with probabilities
proportional to the weights $W_i$, using reproducible systematic resampling.
Repeated draws of the same stored sample are evolved only once, while
retaining their multiplicities. For each selected physical parameter set,
we solve the same background equations and reconstruct $w_{\rm DE}$ from
Eq.~(\ref{wde}). The resulting history $w_{\rm DE}({\cal N})$ is mapped to
redshift using $z=e^{-{\cal N}}-1$ before interpolation, thereby preserving
the correspondence between each redshift and equation-of-state value.
For the pointwise posterior reconstruction shown in the figures below,
these histories are evaluated at 301 redshifts over $0\leq z\leq3$.

At each redshift, we compute the posterior median and the central credible intervals
\begin{equation}
[q_{0.16}(z),q_{0.84}(z)],\qquad
[q_{0.025}(z),q_{0.975}(z)].
\label{pointwisebands}
\end{equation}
The labels $1\sigma$ and $2\sigma$ used in the figures denote the central
68\% and 95\% credible intervals, respectively, rather than the exact
Gaussian probabilities of 68.27\% and 95.45\%. These intervals are
pointwise in redshift and should not be interpreted as simultaneous
credible regions for the entire function. 
The pointwise median curve generally does not correspond to the solution
evaluated at the posterior mean parameters and need not coincide with any
individual posterior history.
Likewise, its intersection with $w_{\rm DE}=-1$ does not
represent the posterior median of the crossing redshift. We also monitor
failed history evaluations and test the sensitivity of the reconstructed
bands to the resampling size and redshift resolution. Increasing the
number of resampled histories improves the numerical representation of
the posterior bands but does not provide additional independent MCMC
information.

%%%%%%%%%%%%%%%%%%%%%%%%%%%%%%%%%%%%%%
\section{Cosmological constraints}
\label{constraintsec}
%%%%%%%%%%%%%%%%%%%%%%%%%%%%%%%%%%%%%%

%
\subsection{Constraints without RSD}
\label{backgroundresults}

\subsubsection{Background constraints and phantom-divide crossing}

We first consider the baseline combination of DES Year~5 SN~Ia, 
DESI DR2 BAO, and compressed CMB data, without RSD information. 
The six-dimensional background parameter space is sampled directly 
using the physical priors in Table~\ref{tab:svt_priors}. 
The four production chains yield the final convergence diagnostics
\begin{equation}
\Delta_{\rm mean}=0.008676,\qquad
\Delta_{\rm bound}=0.047926,
\label{finalconvergence}
\end{equation}
with a mean acceptance fraction of 0.122. Both stopping criteria in
Eq.~(\ref{stoppingcriteria}) are therefore satisfied. The sampler discards
the first 1000 accepted points of each chain as burn-in, and all retained
samples are analyzed with their original posterior weights as described in
Sec.~\ref{posteriormethod}.

%%%%%%%%%%%%%%%%%%%%%%%%%%%%%%%%
\begin{table}[t]
 \centering
 \caption{Marginalized posterior means and central 68\% credible intervals for the physical background parameters and selected derived quantities obtained from DES Year 5 SN~Ia, DESI DR2 BAO, and compressed CMB data without RSD. The Hubble constant $H_0$ is given in 
 ${\rm km\,s^{-1}\,Mpc^{-1}}$, while all other quantities are dimensionless. The quoted intervals correspond to the weighted $[q_{0.16},q_{0.84}]$ endpoints and are not symmetric errors inferred from 
the plotted distributions.}
 \label{tab:svt_constraints}
 \begingroup
 \renewcommand{\arraystretch}{1.18}
 \begin{ruledtabular}
 \begin{tabular}{lcc}
 Parameter & Posterior mean & 68\% credible interval\\
 \hline
 $H_0$ & $67.0736$ & $[66.5095,\,67.6334]$ \\
 $\Omega_{b0}$ & $0.0497458$ & $[0.0488816,\,0.0506173]$ \\
 $\Omega_{m0}$ & $0.311773$ & $[0.306534,\,0.316990]$ \\
 $f_\chi$ & $0.421421$ & $[0.183001,\,0.649540]$ \\
 $s_\chi$ & $0.163257$ & $[0.045418,\,0.269184]$ \\
 $\lambda$ & $1.797894$ & $[1.146897,\,2.458537]$ \\
 \hline
 $\Omega_{{\rm DE},0}$ & $0.688134$ & $[0.682916,\,0.693374]$ \\
 $\Omega_{\chi0}$ & $0.290149$ & $[0.125719,\,0.447198]$ \\
 $s_\chi\Omega_{{\rm DE},0}$ & $0.112401$ & $[0.031232,\,0.185447]$ \\
 $w_{{\rm DE},0}$ & $-0.916457$ & $[-0.943556,\,-0.884263]$ \\
 \end{tabular}
 \end{ruledtabular}
 \endgroup
\end{table}
%%%%%%%%%%%%%%%%%%%%%%%%%%%%%

%%%%%%%%%%%%%%%%%%%%%%%%%%%%%
\begin{figure*}[t]
 \centering
 \includegraphics[width=0.94\textwidth,height=0.77\textheight,keepaspectratio]{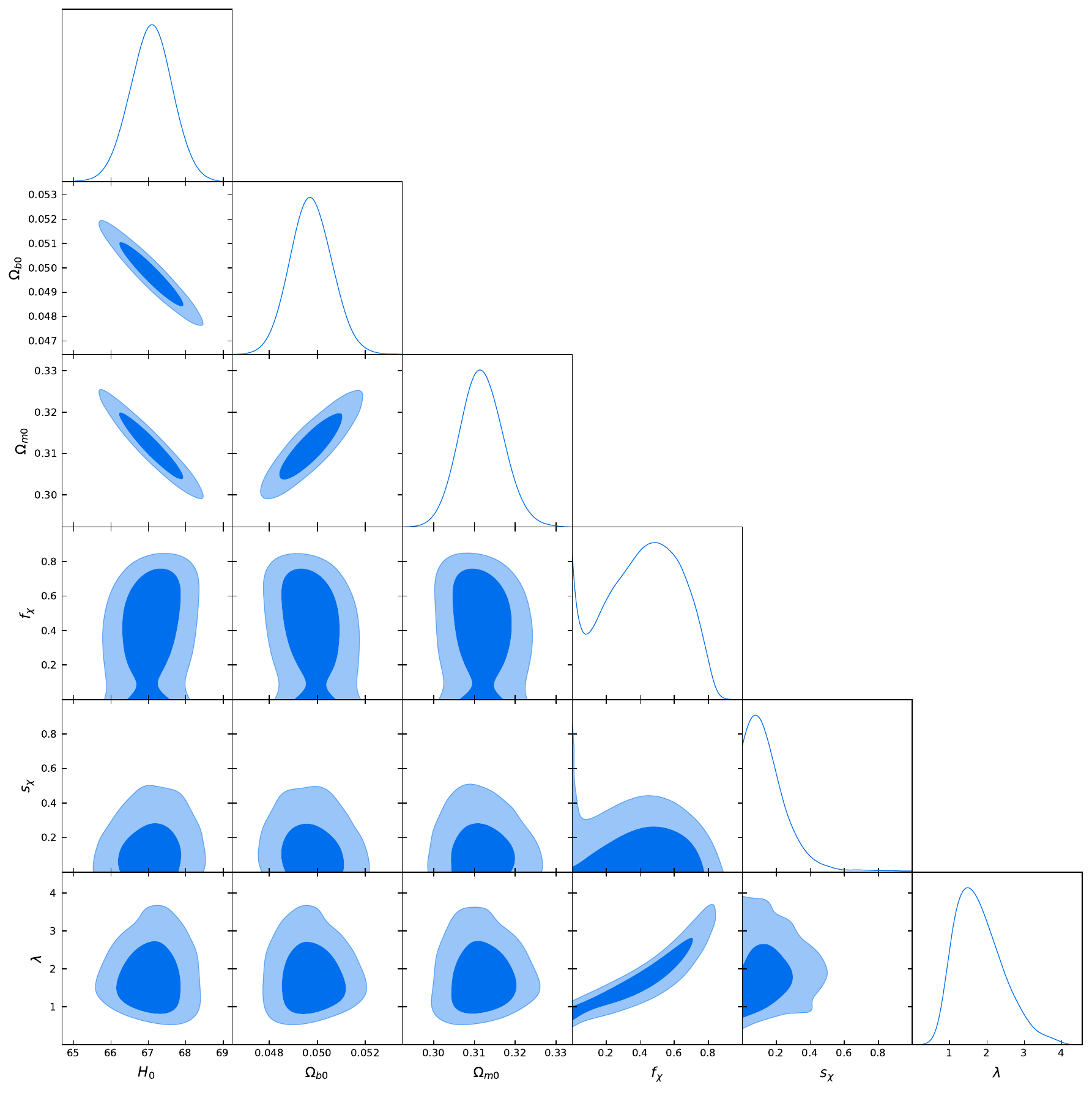}
 \caption{Background-only constraints on
$(H_0,\Omega_{b0},\Omega_{m0},f_\chi,s_\chi,\lambda)$ from
DES Year 5 SN~Ia, DESI DR2 BAO, and compressed CMB data. Diagonal panels show the
one-dimensional marginalized posterior densities; dark- and light-blue contours show the
central 68\% and 95\% two-dimensional marginalized credible regions, respectively.
The Hubble constant $H_0$ is given in
${\rm km\,s^{-1}\,Mpc^{-1}}$, while all other quantities are dimensionless.
All six parameters are sampled directly with the physical priors of
Table~\ref{tab:svt_priors}. The curved
$f_\chi$--$\lambda$ correlation and the small-$f_\chi$ extension in the
$f_\chi$--$s_\chi$ plane are clearly visible. No RSD information is included.
The contours are obtained from the converged four-chain MCMC production
posterior; the matched PolyChord calculation is used separately for the
Bayesian evidence.}
 \label{fig:backgroundtriangle}
\end{figure*}
%%%%%%%%%%%%%%%%%%%%%%%%%%%%%

Figure~\ref{fig:backgroundtriangle} shows the marginalized posterior distributions 
of the six sampled parameters. The geometrical parameters are well constrained, 
with the main posterior support around $H_0 \simeq 67\,{\rm km\,s^{-1}\,Mpc^{-1}}$,
$\Omega_{b0}\simeq0.050$, and $\Omega_{m0}\simeq0.31$.
The correlations among these parameters mainly reflect the compressed CMB 
constraints on the physical densities $\omega_b$ and $\omega_{bc}$, 
together with the BAO distance scale and the SN~Ia luminosity-distance relation. 
For example, at approximately fixed $\omega_b$, one has 
$\Omega_{b0}\propto H_0^{-2}$. The posterior means and central credible intervals are given in Table~\ref{tab:svt_constraints}. The rounded values quoted here serve only to 
indicate the locations of the contours.

The DE composition parameters exhibit more pronounced non-Gaussian features.
The marginalized posterior of $f_\chi$ has broad support at intermediate vector
fractions together with a smaller low-$f_\chi$ feature near the lower prior
boundary. The $s_\chi$ distribution is concentrated at small positive values and
decreases toward its upper prior boundary, whereas $\lambda$ is broadly supported 
over values of order unity. The support for order-unity
$\lambda$, well away from the lower prior boundary, is consistent with
appreciable scalar-field evolution along the potential, as required for the
crossing mechanism described in Sec.~\ref{backgroundtheory}. In the $f_\chi$--$\lambda$ plane, the two-dimensional posterior forms a
single connected, curved band with a clear positive correlation: larger
$f_\chi$ is associated with larger $\lambda$. This behavior can be understood
qualitatively because increasing the present vector fraction reduces the scalar
contribution, $\Omega_{\phi0}=(1-f_\chi)\Omega_{{\rm DE},0}$, while changing
$\lambda$ alters the scalar-field evolution. Correlated changes in $f_\chi$ and
$\lambda$ can therefore yield similar background distance histories.

The extension toward larger $s_\chi$ at small $f_\chi$ has a separate origin. At positive redshifts, the factor $e^{-{\cal D}}$ in Eqs.~(\ref{algebraicOchi}) and (\ref{algebraicQchi}) can strongly suppress the vector contribution even when $s_\chi$ is not very small. Consequently, a rapid change confined to low redshifts can have only a limited impact on integrated distance observables. The matter-era approximation $w_{\rm DE}\simeq-1-s$ is therefore not generally applicable to the present scalar--vector system, since it assumes that the vector field dominates the DE sector. Both the scalar contribution and the evolving vector fraction must be taken into account when interpreting the allowed parameter region.

%%%%%%%%%%%%%%%%%%%%%%%%%%%%%%%%%%%%
\begin{figure}[ht]
 \centering
 \includegraphics[width=\columnwidth]{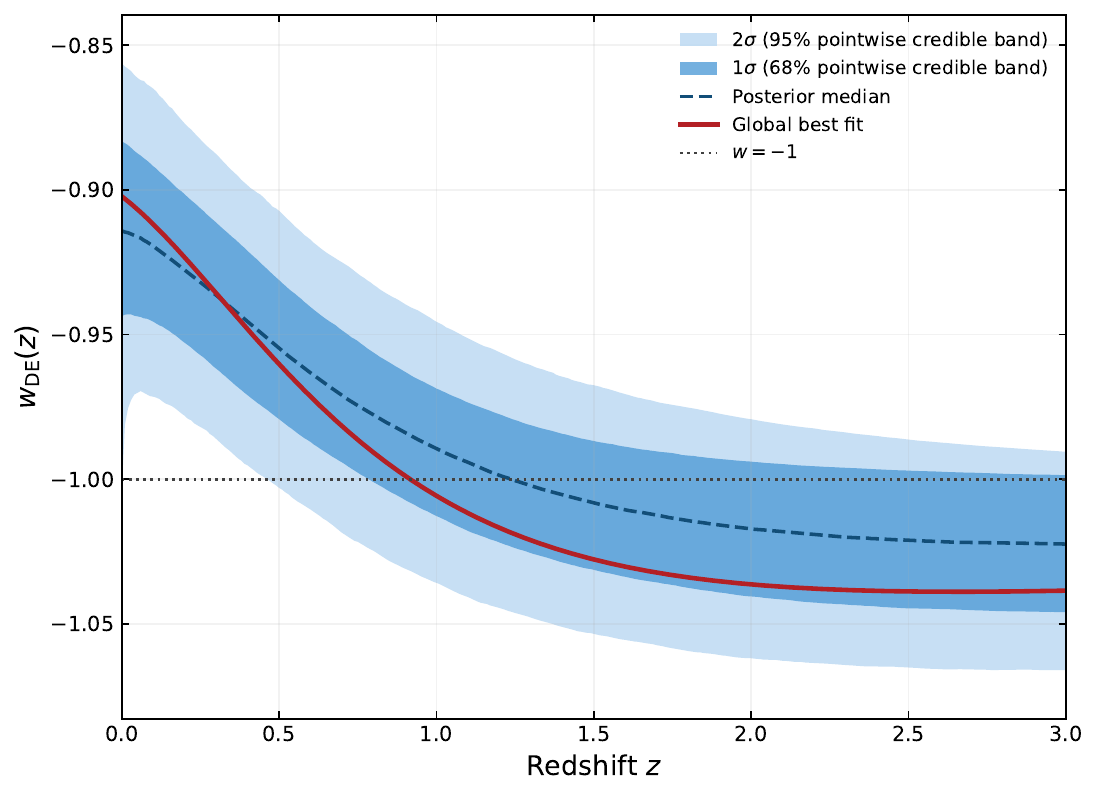}
 \caption{
Reconstruction of $w_{\rm DE}(z)$ for the same no-RSD data and priors as
Fig.~\ref{fig:backgroundtriangle}.  The solid red curve shows the global
best-fit sampled history, while the dashed dark-blue curve denotes the
pointwise posterior median.  The dark- and light-blue bands show the central
pointwise 68\% and 95\% credible intervals, respectively, and the black
dotted line marks $w_{\rm DE}=-1$.  Crossings of individual summary curves
do not by themselves determine the posterior probability for a complete
history to cross the phantom divide.  The rapid variation of the lower
95\% boundary near $z=0$ is discussed in the text.  The figure shows
$0\leq z\leq3$, whereas the directional crossing probability quoted below
is evaluated independently over $0<z\leq4$ using the weighted PolyChord
posterior histories.
}
 \label{fig:backgroundwde}
\end{figure}
%%%%%%%%%%%%%%%%%%%%%%%%%%%%%%%%%%%%%%

The reconstructed DE equation of state is shown in
Fig.~\ref{fig:backgroundwde}, without assuming a phenomenological
$w_0$--$w_a$ parametrization. With $w_{{\rm DE},0}$ denoting the
present-day value of $w_{\rm DE}$, the global best-fit
trajectory has $w_{{\rm DE},0}\simeq-0.90$ and crosses the phantom
divide near $z\simeq0.9$. The pointwise median has
$w_{{\rm DE},0}\simeq-0.91$ and crosses the phantom divide at
$z\simeq1.2$--$1.3$. At $z=2.5$, the corresponding values of
$w_{\rm DE}$ for the global best-fit and pointwise-median curves are
approximately $-1.04$ and $-1.02$, respectively. Thus, both curves
exhibit an evolution from $w_{\rm DE}<-1$ at higher redshifts to
$w_{\rm DE}>-1$ at the present epoch. The difference between these two
summary curves reflects the curved and non-Gaussian posterior structure
shown in Fig.~\ref{fig:backgroundtriangle}.

The crossing mechanism can be understood directly from Eq.~(\ref{wde}).
On the expanding branch with $\epsilon_H<0$, the vector contribution
$\epsilon_H Q_\chi$ is negative, whereas the canonical-scalar contribution
$3x^2$ is nonnegative.  The phantom-divide crossing occurs at the redshift
$z_\times$ satisfying
\begin{equation}
3x^2(z_\times)=-\epsilon_H(z_\times)Q_\chi(z_\times).
\label{crossingcondition}
\end{equation}
Thus, the DE equation of state can cross $w_{\rm DE}=-1$ while the
canonical scalar field retains a positive kinetic term.  This background
mechanism, however, does not by itself guarantee perturbative stability,
which is separately imposed in the RSD analysis of
Sec.~\ref{rsdresults}.

The shaded bands should be interpreted pointwise.  At $z=0$, the lower edge
of the central 95\% credible interval lies above $w_{\rm DE}=-1$, whereas
$w_{\rm DE}=-1$ falls within the allowed bands at positive redshift and
remains consistent with the posterior over much of the plotted range.  The
posterior mean,
$w_{{\rm DE},0}=-0.916457$, together with the central 68\% interval
$[-0.943556,-0.884263]$, also favors $w_{{\rm DE},0}>-1$ for this model and
data combination.  A local Gaussian approximation to the one-dimensional
posterior would place $w_{{\rm DE},0}=-1$ about $2.8\sigma$ from the mean.
This number, however, does not quantify the significance of phantom-divide
crossing, nor does the crossing of the median curve, since crossing is a
property of an individual background history.  We therefore define a
directional crossing by
\begin{eqnarray}
& &
 w_{\rm DE}(0)>-1,\qquad
 w_{\rm DE}(z)<-1 \nonumber \\
& &
 \hbox{for at least one}\quad 0<z\leq4.
 \label{directionalcrossingcriterion}
\end{eqnarray}

The corresponding crossing probability is obtained directly from the
weighted fraction of posterior histories satisfying
Eq.~(\ref{directionalcrossingcriterion}), with all noncrossing histories
retained in the denominator.  Introducing an indicator variable 
$I_i^{\rm cross}$, defined by
$I_i^{\rm cross}=1$ if the $i$th posterior history satisfies
Eq.~(\ref{directionalcrossingcriterion}) and
$I_i^{\rm cross}=0$ otherwise, we compute
\begin{equation}
 P_{\rm cross}=\frac{\sum_i w_i I_i^{\rm cross}}{\sum_i w_i},
 \label{crossingprobabilitydefinition}
\end{equation}
where $w_i$ denotes the original nested-sampling posterior weight.  For the
matched DES-SN5YR+DESI DR2 BAO+compressed-CMB posterior, we obtain
\begin{equation}
 P_{\rm cross}^{\rm no\mbox{-}RSD}=0.791276.
 \label{pcrossnorsd}
\end{equation}
Thus, about $79.1\%$ of the posterior weight corresponds to histories with
$w_{\rm DE}(0)>-1$ and $w_{\rm DE}(z)<-1$ at some redshift
$0<z\leq4$.  As numerical checks, systematic resampling to 4000
equal-weight histories gives $P_{\rm cross}=0.791750$, while increasing the
reconstruction grid from 301 to 601 points over $0\leq z\leq4$ leaves the
direct weighted result unchanged to the quoted precision.

The sharp variation of the lower 95\% boundary near $z=0$ should likewise
not be overinterpreted. For small $f_\chi$, the suppression exponent has a
characteristic transition scale
$\Delta\ell\sim f_\chi/(2s_\chi)$, where
$\ell=\ln E=\ln(H/H_0)$. Thus, a finite positive $s_\chi$ can produce a
narrow feature close to the present epoch while the vector contribution
remains strongly suppressed at higher redshifts. The visual appearance of
this feature can also be affected by interpolating the 6000-point background
solution onto the 301-point plotting grid and by the precise endpoint
normalization. We therefore interpret Fig.~\ref{fig:backgroundwde} as
showing that smooth crossing histories are allowed by the posterior, rather
than as evidence for a detected sharp transition at very low redshift.

\subsubsection{Matched comparison with flat $\Lambda$CDM}

We next ask whether the improved background fit of the SVT model is sufficient
to compensate for its three additional parameters relative to flat
$\Lambda$CDM.  To make this comparison, we use exactly the same DES-SN5YR,
DESI DR2 BAO, and compressed-CMB likelihoods for the SVT and flat
$\Lambda$CDM models, with no RSD data included.  The flat $\Lambda$CDM
analysis samples $(H_0,\Omega_{b0},\Omega_{m0})$ with the same priors as the
corresponding SVT parameters, and the supernova absolute magnitude is
marginalized in the same way in both models.  For each model, the quoted best
fit is the stored MCMC point minimizing
$\chi^2_{\rm back}=\chi^2_{\rm SN}+\chi^2_{\rm BAO}+\chi^2_{\rm CMB}$.
Throughout this section, model differences are defined as SVT minus flat
$\Lambda$CDM,
\begin{equation}
 \Delta\chi^2_{\rm min}=\chi^2_{\rm min,SVT}-
 \chi^2_{\rm min,\Lambda{\rm CDM}}\,,
 \label{deltachi2definition}
\end{equation}
so that negative values correspond to a better best fit for SVT.  We also
quote the Akaike information criterion (AIC) \cite{Akaike:1974},
$\mathrm{AIC}=\chi^2_{\rm min}+2k$, and the Bayesian information criterion
(BIC) \cite{Schwarz:1978}, $\mathrm{BIC}=\chi^2_{\rm min}+k\ln N$, where
$k$ denotes the number of sampled cosmological parameters used in each model.
Thus $k_{\rm SVT}=6$ and $k_{\Lambda{\rm CDM}}=3$ for the no-RSD comparison.
We adopt the
conventional count $N=1845$, comprising 1829 supernova entries, 13 BAO
observables, and three compressed-CMB observables.

%%%%%%%%%%%%%%%%%%%%%%%%
\begin{table*}[t]
 \centering
 \caption{Matched no-RSD model comparison using
 DES-SN5YR+DESI DR2 BAO+compressed CMB.  The final row gives SVT minus
 $\Lambda$CDM.  The best-fit entries are stored MCMC points.  The
 $\ln Z_{\rm NS}^{\rm I}$ column contains the final Cobaya-corrected matched
 PolyChord log evidences under Convention~I (the rectangular valid-volume
 convention, defined below).  In the final row, the entry in this column is
 $\Delta\ln B_{\rm NS}^{\rm I}$.  Convention~II is not tabulated; the
 corresponding Bayes factor is derived below in the main text.  The quoted
 evidence errors are PolyChord numerical uncertainties, and the uncertainty
 on $\Delta\ln B_{\rm NS}^{\rm I}$ is propagated in quadrature.}
 \label{tab:modelcomparison}
 \begin{ruledtabular}
 \begin{tabular}{lrrrrrrrr}
 Model & $k$ & $\chi^2_{\rm SN}$ & $\chi^2_{\rm BAO}$ &
 $\chi^2_{\rm CMB}$ & $\chi^2_{\rm min}$ & AIC & BIC & $\ln Z_{\rm NS}^{\rm I}$\\
 \hline
 $\Lambda$CDM & 3 & 1649.580 & 12.366 & 0.942 &
 1662.889 & 1668.889 & 1685.449 & $-844.366\pm0.101$\\
 SVT & 6 & 1639.152 & 9.513 & 2.120 &
 1650.785 & 1662.785 & 1695.906 & $-843.646\pm0.128$\\
 \hline
 SVT$-\Lambda{\rm CDM}$ & $+3$ & $-10.428$ & $-2.853$ & $+1.178$ &
 $-12.104$ & $-6.104$ & $+10.457$ & $+0.720\pm0.163$\\
 \end{tabular}
 \end{ruledtabular}
\end{table*}
%%%%%%%%%%%%%%%%%%%%%%%%

The best-fit stored samples give $\chi^2_{\rm min}=1650.785$ for SVT and
$\chi^2_{\rm min}=1662.889$ for flat $\Lambda$CDM, corresponding to
$\Delta\chi^2_{\rm min}=-12.104$.  Most of this improvement comes from the
supernova likelihood, with $\Delta\chi^2_{\rm SN}=-10.428$, while the BAO
contribution, $\Delta\chi^2_{\rm BAO}=-2.853$, provides a further improvement
that is partly offset by $\Delta\chi^2_{\rm CMB}=+1.178$.  These values are
evaluated at the best stored chain points and do not result from a separate
continuous minimization.  We do not translate $\Delta\chi^2_{\rm min}$ into
an equivalent Gaussian significance, because flat $\Lambda$CDM lies on a
boundary of the SVT parameter space and some of the additional SVT parameters
become non-identifiable in this limit.  The regularity conditions required
for the standard application of Wilks' theorem are therefore not satisfied.

The AIC difference between SVT and flat $\Lambda$CDM is
\begin{equation}
 \Delta\mathrm{AIC}=-6.104.
 \label{deltaaicnorsd}
\end{equation}
The corresponding BIC difference is $\Delta\mathrm{BIC}=+10.457$.  Thus,
the improvement in the SVT best fit is sufficient to overcome the
AIC parameter penalty $2(k_{\rm SVT}-k_{\Lambda{\rm CDM}})=6$, whereas
the larger BIC penalty,
$(k_{\rm SVT}-k_{\Lambda{\rm CDM}})\ln N=3\ln(1845)$, favors the
simpler flat $\Lambda$CDM model within the BIC approximation.  We do not interpret the BIC difference as an approximation to the Bayes factor in the present analysis.  The standard
BIC approximation assumes a regular, locally Gaussian posterior around an
identifiable interior maximum and also depends on the adopted effective
number of data points.  These assumptions are not well satisfied here,
because the SVT posterior exhibits curved degeneracies and the flat
$\Lambda$CDM limit is a nonregular boundary at which additional SVT
parameters become non-identifiable.  
We therefore adopt the directly computed Bayesian evidence for model
comparison under the specified prior measure.

We evaluate the Bayesian evidence using matched
PolyChord~\cite{Handley:2015polychord} production runs for SVT and flat
$\Lambda$CDM, with identical likelihood implementations and normalized
top-hat priors for the sampled parameters.  Each run is configured with
500 live points, $n_{\rm prior}=15000$, and a PolyChord precision criterion
of 0.01.  Here $n_{\rm prior}$ is the number of valid initial-prior points
used by the Cobaya PolyChord implementation to estimate the fraction of the
nominal rectangular prior volume that survives the imposed validity
conditions.  We denote this fraction by $f_{\rm valid}$.  If
$n_{\rm discarded}$ denotes the total number of prior trials required to
obtain these $n_{\rm prior}$ valid points, its Monte Carlo estimate is
\begin{equation}
 f_{\rm valid}=\frac{n_{\rm prior}}{n_{\rm discarded}}.
 \label{validpriorfraction}
\end{equation}
We denote by $Z_{\rm raw}$ the evidence returned directly by PolyChord before
this prior-volume correction.  The corrected evidence then satisfies
\begin{equation}
 \ln Z=\ln Z_{\rm raw}+\ln f_{\rm valid}.
 \label{evidencevolumecorrection}
\end{equation}
For SVT, the raw nested-sampling result is
$\ln Z_{\rm raw}=-838.52996\pm0.12757$.  The initial-prior calculation gives
$n_{\rm discarded}=2499013$, corresponding to
$\ln f_{\rm valid}=-5.11560$, and hence, under Convention~I,
$\ln Z_{\rm SVT,noRSD}^{\rm I}=-843.64556\pm0.12757$.  For flat $\Lambda$CDM, the
corresponding values are
$\ln Z_{\rm raw}=-840.35348\pm0.10111$ and
$n_{\rm discarded}=828954$, giving
$\ln f_{\rm valid}=-4.01211$ and
$\ln Z_{\Lambda{\rm CDM,noRSD}}^{\rm I}=-844.36559\pm0.10111$.

The compressed-CMB likelihood is restricted to a common early-time
calibration domain occupying the fraction
$f_{\rm early}=0.0178433642$ of the rectangular prior volume, corresponding
to $\ln f_{\rm early}=-4.02612$.  For flat $\Lambda$CDM, this restriction
accounts essentially for the entire reduction of the valid prior volume:
the value of $f_{\rm valid}$ inferred from the initial-prior sampling is
consistent with $f_{\rm early}$ within its finite Monte Carlo uncertainty.
For SVT, an additional restriction arises because only a subset of points
within the calibration domain admits a regular radiation-era shooting
solution.  The production calculation gives
\begin{equation}
 \frac{f_{\rm valid}}{f_{\rm early}}=0.33639,
 \label{svtrootfractionproduction}
\end{equation}
so that only about $33.6\%$ of the calibrated SVT prior volume lies on the
regular shooting branch.  Defining $f_{\rm root}$ as this conditional
fraction, the SVT valid volume can be written as
$f_{\rm valid}=f_{\rm early}f_{\rm root}$.

We determined $f_{\rm root}$ independently with a dedicated prior-only
Monte Carlo calculation.  Uniform samples were drawn from the same
rectangular parameter priors and restricted to the common compressed-CMB
calibration domain.  For each conditioned point, we applied the same
radiation-era shooting procedure and forward background evolution as in the
production analysis, without evaluating the cosmological likelihood, and
recorded whether a regular physical solution existed. 
Of $50000$ conditioned points, $16821$ yielded regular solutions, 
with no numerical exceptions.
We therefore obtain
\begin{align}
 f_{\rm root}^{\rm no\mbox{-}RSD}&=0.33642\pm0.00211,\\
 \ln f_{\rm root}^{\rm no\mbox{-}RSD}&=-1.08939\pm0.00628.
\end{align}
This independent estimate agrees closely with the production value
$f_{\rm valid}/f_{\rm early}=0.33639$.  The agreement shows that the
additional reduction of the SVT prior volume is associated with the
existence condition for the regular radiation-era shooting branch, rather
than with numerical failures or an unaccounted-for likelihood cut.

We quote the Bayesian comparison under two explicit prior conventions.
In Convention~I (the rectangular valid-volume convention), the sampled
top-hat box defines the prior measure, and the prior-volume penalty is retained
for points at which the forward model does not yield a regular solution.
In Convention~II (the regular-physical-support convention), the SVT prior is
instead conditioned on the regular radiation-era branch, so that the
existence of this branch is treated as part of the definition of the physical
cosmological model rather than as an Occam penalty.  The common
compressed-CMB calibration-domain condition is imposed on both models and
therefore cancels in the matched Bayes factor.

We define
$\Delta\ln B_{\rm NS}^{X}
=\ln Z_{\rm SVT,noRSD}^{X}
-\ln Z_{\Lambda{\rm CDM,noRSD}}^{X}$
and $B_{\rm NS}^{X}=\exp(\Delta\ln B_{\rm NS}^{X})$,
where $X={\rm I},{\rm II}$ denotes the prior convention.  The matched
production results are
\begin{align}
 \ln Z_{\rm SVT,noRSD}^{\rm I}&=-843.646\pm0.128,\nonumber\\
 \ln Z_{\Lambda{\rm CDM,noRSD}}^{\rm I}&=-844.366\pm0.101,\nonumber\\
 \Delta\ln B_{\rm NS}^{\rm I}&=0.720\pm0.163,
 & B_{\rm NS}^{\rm I}&=2.05,\nonumber\\
 \Delta\ln B_{\rm NS}^{\rm II}&=1.809\pm0.163,
 & B_{\rm NS}^{\rm II}&=6.11.
 \label{bayesfactorresult}
\end{align}
No new PolyChord run is required to pass from Convention~I to Convention~II.
Using the independently measured regular-branch fraction above, we obtain
\begin{align}
 \Delta\ln B_{\rm NS}^{\rm II}
 &=\Delta\ln B_{\rm NS}^{\rm I}
   -\ln f_{\rm root}^{\rm no\mbox{-}RSD}\nonumber\\
 &=1.80942\pm0.16290.
\end{align}
The quoted uncertainty is obtained by adding in quadrature the PolyChord
evidence uncertainty and the binomial Monte Carlo uncertainty associated
with the independently measured regular-branch fraction.  The latter
corresponds to
$\sigma(\ln f_{\rm root})=0.00628$, which is much smaller than the
nested-sampling contribution and therefore has a negligible effect on the
total uncertainty.

Thus, the no-RSD background data yield
$\Delta\ln B_{\rm NS}>0$ under both prior conventions, corresponding to
$B_{\rm NS}>1$ in favor of SVT, with a smaller value under Convention~I
and a larger value under Convention~II.  The two results are based on the
same likelihood and posterior distribution over the regular branch; they
differ only in the prior normalization assigned to the regular physical
support.

As an independent numerical robustness check, a separate matched PolyChord
calculation with $n_{\rm live}=400$, analyzed with the same Convention~I
normalization, gives
$\ln Z_{\rm SVT,noRSD}^{\rm I}=-843.460\pm0.150$ and
$\ln Z_{\Lambda{\rm CDM,noRSD}}^{\rm I}=-844.448\pm0.124$, corresponding to
$\Delta\ln B_{\rm NS}^{\rm I}=0.988\pm0.194$.  The difference of 0.268 from
the Convention~I production value in Eq.~(\ref{bayesfactorresult}) is about
$1.1\sigma$ after combining the quoted numerical uncertainties.  The two
calculations therefore agree at the level expected from their evidence
errors, and both yield $\Delta\ln B_{\rm NS}^{\rm I}>0$.  Together with the
agreement of the independently measured regular-root fraction quoted above,
this provides a useful check on both the nested-sampling calculation and the
prior-volume accounting.

The distinction between the two conventions concerns the normalization of
the prior support, not the posterior sampling on the regular branch.
Switching between them changes the absolute SVT evidence normalization and
the Bayes factor, but it does not alter the posterior constraints,
reconstructed $w_{\rm DE}(z)$ histories, best-fit $\chi^2$, AIC, or BIC
quoted above.  We therefore report both Bayes factors explicitly.

The no-RSD analysis therefore provides a coherent background-level picture. The posterior assigns about $79.1\%$ 
of its weight to regular histories with
$w_{\rm DE}(0)>-1$ and $w_{\rm DE}(z)<-1$ 
at some redshift $0<z\leq4$.
The improved SVT best fit is sufficient to overcome the additional-parameter penalty in the AIC, whereas the BIC penalizes the larger parameter space more
strongly, and the matched Bayesian evidence gives
$\Delta\ln B_{\rm NS}>0$ under both prior conventions considered above.
The RSD analysis in Sec.~\ref{rsdresults} tests whether this background picture persists after growth information is added, while also imposing the perturbative support conditions, and provides a second matched evidence
comparison under the same two prior conventions.

%%%%%%%%%%%%%%%%%%%%%%%%%%%%%%%%%%%%%%
\subsection{Constraints including RSD}
\label{rsdresults}
%%%%%%%%%%%%%%%%%%%%%%%%%%%%%%%%%%%%%%

\subsubsection{RSD constraints and phantom-divide crossing}

We now augment the DES-SN5YR+DESI DR2 BAO+compressed-CMB baseline of
Sec.~\ref{backgroundresults} with the Gold-2017 RSD likelihood in
Eq.~(\ref{rsdlikelihood}).  The sampled space is enlarged from
$\boldsymbol\vartheta_{\rm bg}$ to $\boldsymbol\vartheta_{\rm RSD}$, and the
growth equation and theoretical support conditions are those summarized in
Sec.~\ref{perturbationtheory}, while the approximate fiducial-cosmology
geometry correction and the RSD covariance are specified in
Sec.~\ref{rsdmethod}. The direct-coordinate RSD production chains employ the
learned full nine-dimensional covariance. Their final convergence diagnostics
are
\begin{equation}
 \Delta_{\rm mean}=0.009227,\qquad
 \Delta_{\rm bound}=0.046972,
 \label{finalconvergencersd}
\end{equation}
with a mean acceptance fraction of 0.496. Thus both stopping conditions in
Eq.~(\ref{stoppingcriteria}) are satisfied.

\begin{figure*}[t]
 \centering
 \includegraphics[width=0.94\textwidth,height=0.80\textheight,keepaspectratio]{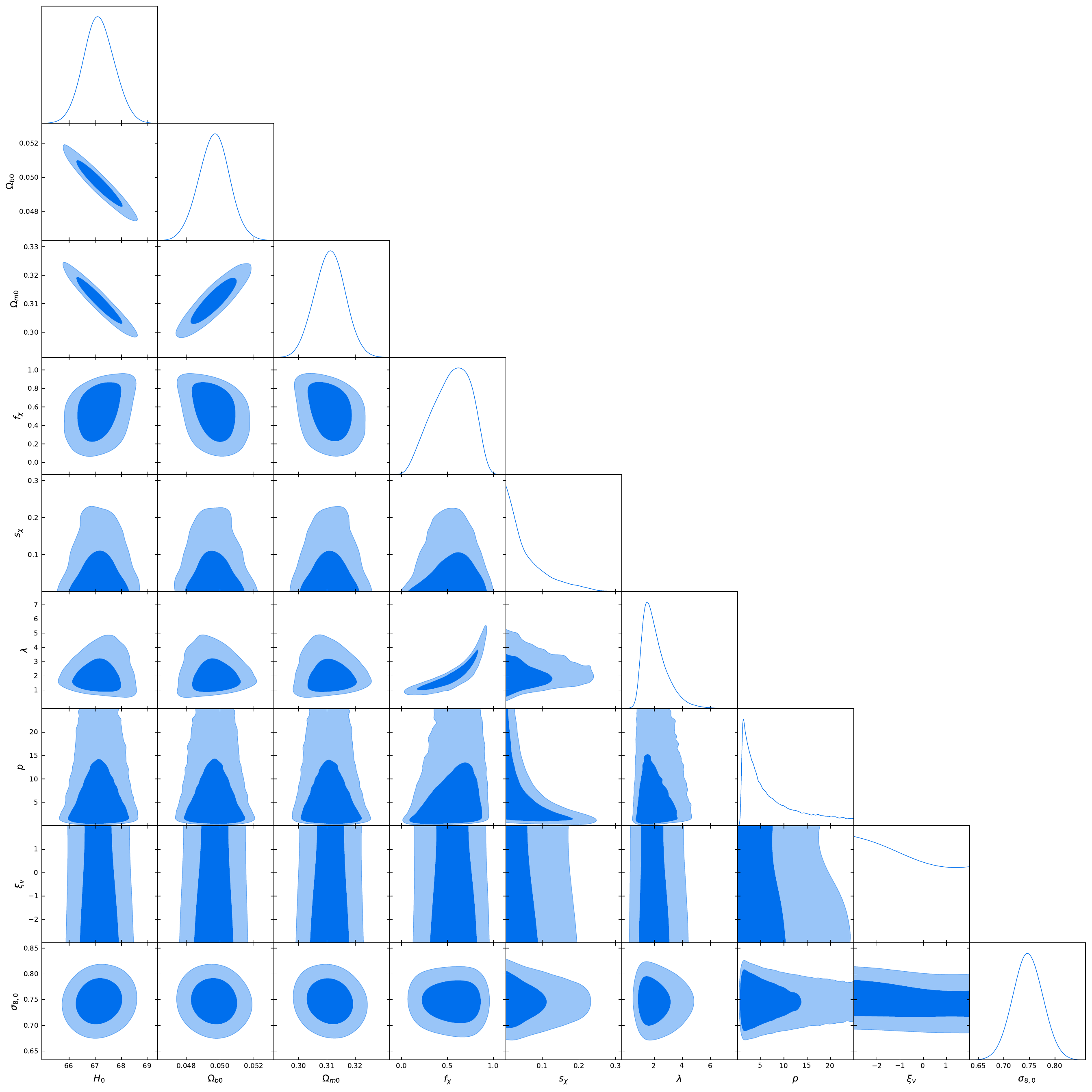}
 \caption{Marginalized posterior constraints on
$(H_0,\Omega_{b0},\Omega_{m0},f_\chi,s_\chi,\lambda,p,
\log_{10}\nu_v,\sigma_{8,0})$ from
DES Year 5 SN~Ia, DESI DR2 BAO, compressed CMB, and Gold-2017 RSD data.
The plotting conventions are the same as in
Fig.~\ref{fig:backgroundtriangle}.  All nine parameters are sampled jointly
with the priors in Table~\ref{tab:svt_priors}, subject to the perturbative
support conditions described in Sec.~\ref{perturbationtheory}.  The RSD data
constrain $\sigma_{8,0}$ and the growth history, while the posteriors for
$p$ and $\log_{10}\nu_v$ remain broad and non-Gaussian.  The contours are
obtained from the converged four-chain MCMC production posterior; PolyChord
is used separately for the Bayesian-evidence calculation.}
 \label{fig:rsdtriangle}
\end{figure*}

Figure~\ref{fig:rsdtriangle} shows that the common background parameters
remain concentrated near
$H_0\simeq67\,{\rm km\,s^{-1}\,Mpc^{-1}}$,
$\Omega_{b0}\simeq0.050$, and $\Omega_{m0}\simeq0.31$, with locations and
correlation patterns similar to those in the no-RSD analysis shown in
Fig.~\ref{fig:backgroundtriangle}.  This behavior is consistent with the
strong constraints already imposed on the background cosmological parameters
by the compressed CMB and BAO data.  The fluctuation-amplitude parameter is
constrained to $\sigma_{8,0}\simeq0.75$.  Its correlations with the
background parameters are relatively weak in the displayed projections,
whereas its correlations with the perturbation-sector parameters arise
because the RSD observable $f\sigma_8(z)$ depends on both the growth history
and its present-day normalization.

A notable difference from the no-RSD posterior is that the secondary
small-$f_\chi$ peak visible in Fig.~\ref{fig:backgroundtriangle} is absent in
Fig.~\ref{fig:rsdtriangle}.  This change cannot be attributed to the RSD
likelihood alone, because the RSD analysis simultaneously imposes the
perturbative support conditions that are absent from the no-RSD baseline.
In particular, the condition $ps\leq1$, together with
$s=s_\chi/f_\chi$, requires
\begin{equation}
 p s_\chi\leq f_\chi.
 \label{smallfchisupport}
\end{equation}
Hence, as $f_\chi$ approaches zero, configurations with finite $s_\chi$ are
strongly restricted.  This removes much of the small-$f_\chi$ extension
present in the background-only posterior and correspondingly suppresses the
marginal posterior volume associated with the secondary peak.  The
positivity condition on $c_\psi^2$ and the measured $f\sigma_8(z)$ values can
further restrict the surviving small-$f_\chi$ configurations, but the present 
analysis does not disentangle the effect of the RSD data
from that of the additional theoretical support conditions.

In Fig.~\ref{fig:rsdtriangle}, the posterior for $s_\chi$ is concentrated at
small positive values, while that for $\lambda$ peaks at values of order
unity and decreases toward larger $\lambda$. 
The one-dimensional $p$ and
$\log_{10}\nu_v$ posteriors remain broad and non-Gaussian, so neither is well
represented by a symmetric Gaussian error.  This is physically natural
because the RSD observable constrains a combination of the effective
gravitational coupling, longitudinal-vector sound speed, growth factor, and
$\sigma_{8,0}$ rather than measuring $p$ and $\nu_v$ separately.

More generally, differences between the RSD and no-RSD background
marginals should be interpreted with the enlarged parameter space and the
additional perturbative support in mind, rather than being attributed to the
RSD likelihood alone.  The most direct information added by the growth data
is the constraint on $\sigma_{8,0}$ and the exclusion of histories
inconsistent with the measured $f\sigma_8(z)$ values.

%%%%%%%%%%%%%%%%%%%%%%%%%%%
%%%%%%%%%%%%%%%%%%%%%%

%%%%%%%%%%%%%%%%%%%%%%
\begin{figure}[t]
 \centering
 \includegraphics[width=\columnwidth]{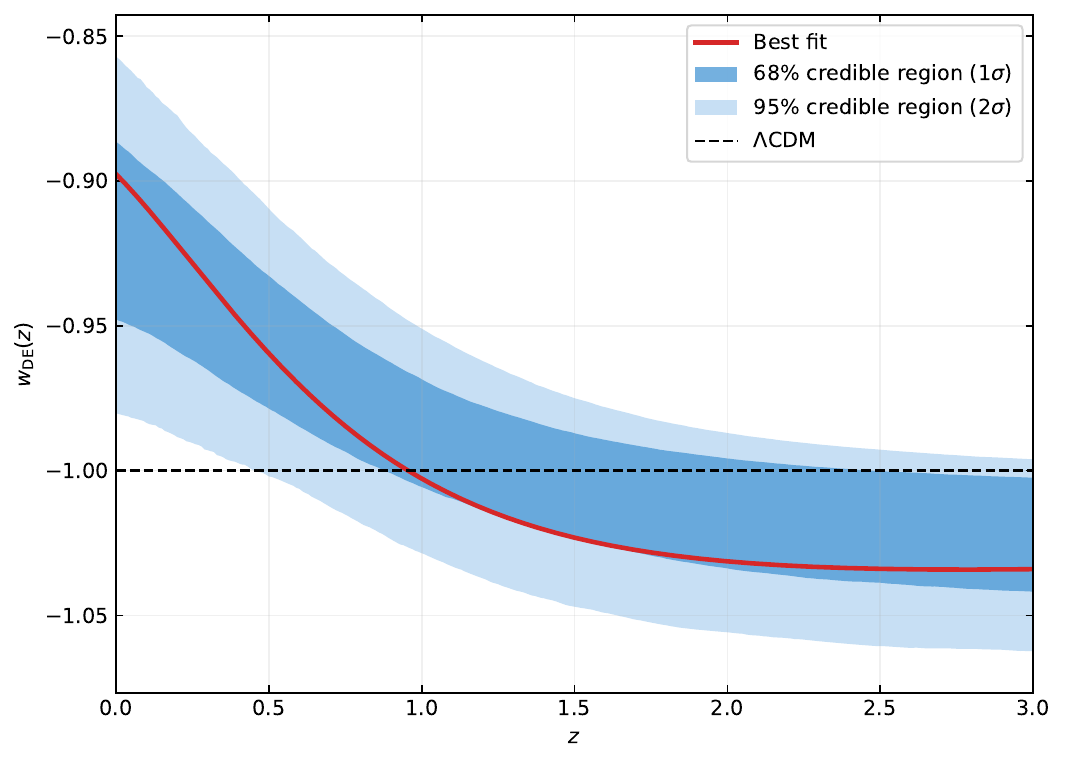}
 \caption{Reconstruction of $w_{\rm DE}(z)$ from the
DES-SN5YR+DESI DR2 BAO+compressed-CMB+RSD posterior in
Fig.~\ref{fig:rsdtriangle}.  The solid red curve shows the global best-fit
stored sample, the dark- and light-blue bands the central pointwise $68\%$ and
$95\%$ credible intervals, and the black dashed line $w_{\rm DE}=-1$.
The interpretation of the pointwise bands is the same as in
Fig.~\ref{fig:backgroundwde}.  The figure covers $0\leq z\leq3$; the
directional crossing probability quoted below is computed from the weighted
PolyChord posterior over $0<z\leq4$.}
 \label{fig:rsdwde}
\end{figure}
%%%%%%%%%%%%%%%%%%%%%%%

The RSD reconstruction of $w_{\rm DE}(z)$ in Fig.~\ref{fig:rsdwde} remains
qualitatively similar to the no-RSD result.  The global best-fit history has
$w_{{\rm DE},0}\simeq-0.90$, crosses $-1$ at approximately $z\simeq0.9$,
and approaches $w_{\rm DE}\simeq-1.03$ at $z\gtrsim2$.  The cosmological-constant value, $w_{\rm DE}=-1$, lies outside the central
pointwise credible bands at $z=0$ but falls within them at positive redshift.  Thus the RSD posterior retains the preference for
$w_{\rm DE}>-1$ at low redshift while allowing histories with
$w_{\rm DE}<-1$ at earlier times.  Compared with
Fig.~\ref{fig:backgroundwde}, the best-fit crossing redshift and the overall
shape change little.  Subject to the support qualification discussed above,
the main additional information in the RSD analysis therefore comes from the
growth sector rather than from a qualitatively different background
expansion history.

Applying the same directional criterion and direct weighted estimator as in
Sec.~\ref{backgroundresults} to the full matched RSD posterior gives
\begin{equation}
 P_{\rm cross}^{\rm RSD}=0.932745.
 \label{pcrossrsd}
\end{equation}
As a numerical check, systematic resampling of the same weighted posterior
into 4000 equal-weight histories gives $P_{\rm cross}=0.932250$, in close
agreement with the direct weighted result.  The latter is also unchanged
when the reconstruction grid is refined from 301 to 601 points over
$0\leq z\leq4$.  The RSD analysis therefore assigns about $93.3\%$ of the
posterior weight to histories satisfying the directional crossing criterion,
compared with $79.1\%$ in the no-RSD baseline.  Since the RSD analysis also
imposes perturbative support conditions that are absent from the no-RSD
baseline, this difference cannot be attributed to the RSD likelihood alone.

\subsubsection{Matched RSD comparison with flat $\Lambda$CDM}

For the RSD model comparison, flat $\Lambda$CDM samples
$(H_0,\Omega_{b0},\Omega_{m0},\sigma_{8,0})$, whereas SVT samples the nine
parameters in Eq.~(\ref{rsdparameters}).  
The minimum $\chi^2$ values among the stored MCMC samples are
$\chi^2_{\rm min,\Lambda{\rm CDM}}=1675.640126$ and
$\chi^2_{\rm min,SVT}=1664.024245$, respectively. 
With the same sign convention as in
Eq.~(\ref{deltachi2definition}), their difference is
\begin{equation}
 \Delta\chi^2_{\rm min,RSD}=-11.615881.
 \label{deltachi2rsd}
\end{equation}
SVT therefore improves the minimum $\chi^2$ by about $11.6$ while introducing
five additional sampled parameters,
$k_{\rm SVT}-k_{\Lambda{\rm CDM}}=9-4=5$.  Using the conventional data-point
count
\begin{equation}
N_{\rm RSD}=1829_{\rm SN}+13_{\rm BAO}+3_{\rm CMB}+18_{\rm RSD}
=1863\,,
\label{nrdsdatacount}
\end{equation}
the corresponding information criteria and Bayesian evidences are summarized
in Table~\ref{tab:modelcomparisonrsd}.

%%%%%%%%%%%%%%%%%%%%%%%%%
\begin{table*}[t]
 \centering
 \caption{Matched RSD model comparison using
 DES-SN5YR+DESI DR2 BAO+compressed CMB+Gold-2017 RSD.  The notation and
 evidence conventions are the same as in Table~\ref{tab:modelcomparison}.
 The $\chi^2_{\rm min}$ values are the best stored MCMC samples, and the final
 row gives SVT minus $\Lambda$CDM.  In that row, the entry in the
 $\ln Z_{\rm NS}^{\rm I}$ column is $\Delta\ln B_{\rm NS,RSD}^{\rm I}$.
 Convention~II is not tabulated and is given below in the main text.}
 \label{tab:modelcomparisonrsd}
 \begin{ruledtabular}
 \begin{tabular}{lrrrrr}
 Model & $k$ & $\chi^2_{\rm min}$ & AIC & BIC & $\ln Z_{\rm NS}^{\rm I}$\\
 \hline
 $\Lambda$CDM & 4 & 1675.640 & 1683.640 & 1705.760 & $-853.161\pm0.118$\\
 SVT & 9 & 1664.024 & 1682.024 & 1731.794 & $-851.514\pm0.133$\\
 \hline
 SVT$-\Lambda$CDM & $+5$ & $-11.616$ & $-1.616$ & $+26.034$ & $+1.647\pm0.178$\\
 \end{tabular}
 \end{ruledtabular}
\end{table*}
%%%%%%%%%%%%%%%%%%%%%%%%%%

Thus, the AIC difference between SVT and flat 
$\Lambda$CDM is
\begin{equation}
 \Delta\mathrm{AIC}=-1.616.
 \label{deltaaicrsd}
\end{equation}
The corresponding BIC difference,
$\Delta \mathrm{BIC} \equiv
\mathrm{BIC}_{\rm SVT}
-\mathrm{BIC}_{\Lambda{\rm CDM}}$,
is $+26.034$.  Thus, the best-fit improvement nearly compensates for the
AIC penalty associated with the 
five additional parameters, whereas the
larger BIC penalty favors the simpler flat 
$\Lambda$CDM model within the
BIC approximation. 
As in
Sec.~\ref{backgroundresults}, we do not identify the BIC difference with a
Bayes-factor approximation and instead use the directly computed
nested-sampling evidence below.

We repeated the matched PolyChord calculation using the same valid-volume
prescription as in Sec.~\ref{backgroundresults}.  Both runs used 500 live
points, $n_{\rm prior}=15000$, and a precision criterion of $0.01$, with
$\texttt{num\_repeats}=45$ for SVT and $20$ for $\Lambda$CDM.  For SVT,
$\ln Z_{\rm raw}=-843.99544\pm0.13309$ and
$n_{\rm discarded}=27633437$ give $\ln f_{\rm valid}=-7.51873$, yielding
\begin{equation}
 \ln Z_{\rm SVT,RSD}^{\rm I}
 =\ln Z_{\rm raw}+\ln f_{\rm valid}
 =-851.51417\pm0.13309.
\end{equation}
For $\Lambda$CDM, the corresponding raw evidence was
$\ln Z_{\rm raw}=-849.13049\pm0.11787$, while
$n_{\rm discarded}=844645$ gives $\ln f_{\rm valid}=-4.03087$ 
and hence
\begin{equation}
 \ln Z_{\Lambda{\rm CDM,RSD}}^{\rm I}
 =-853.16135\pm0.11787.
\end{equation}
The $\Lambda$CDM value of $\ln f_{\rm valid}$ differs by less than $0.005$
from the common early-calibration value
$\ln f_{\rm early}=-4.02612$.  For SVT, the remaining finite-model-volume
correction additionally reflects points rejected by the forward cosmological
solution and by the remaining physical requirements, including
Eq.~(\ref{positivecs2}).  
We adopt the same two prior conventions as in
Sec.~\ref{backgroundresults}.  The analytic perturbative support conditions in
Eq.~(\ref{rsdtheorysupport}) are already incorporated into the RSD sampling
prior and are treated identically in both conventions.  Their
analytic fraction of the original rectangular prior volume is
$f_{\rm stab}=0.0743775165$, corresponding to
$\ln f_{\rm stab}=-2.59860$.

We independently estimated the remaining regular-support fraction with a
second dedicated prior-only MCMC calculation.  The forward shooting calculation
was performed for $50000$ samples drawn from the prior conditioned on both the
common compressed-CMB calibration domain and the analytic stability support.
Of these samples, $20551$ satisfied the remaining regularity and physical
requirements, giving
\begin{align}
 f_{\rm root}^{\rm RSD}&=0.41102\pm0.00220,\\
 \ln f_{\rm root}^{\rm RSD}&=-0.88911\pm0.00535.
\end{align}
No numerical exceptions were encountered.  As a consistency check of the
prior-only sampling, we also measured the fractions of the original
unconditioned prior volume satisfying the analytic stability support and
the common early-calibration condition.  The resulting fractions,
$0.0744065$ and $0.0178571$, respectively, agree closely with the analytic
stability fraction given above and with $f_{\rm early}=0.0178434$.

Using the same Bayes-factor notation and prior conventions as in
Sec.~\ref{backgroundresults}, the matched evidence calculation gives
\begin{align}
 \ln Z_{\rm SVT,RSD}^{\rm I}&=-851.514\pm0.133,\nonumber\\
 \ln Z_{\Lambda{\rm CDM,RSD}}^{\rm I}&=-853.161\pm0.118,\nonumber\\
 \Delta\ln B_{\rm NS,RSD}^{\rm I}&=1.647\pm0.178,
 & B_{\rm NS,RSD}^{\rm I}&=5.19,\nonumber\\
 \Delta\ln B_{\rm NS,RSD}^{\rm II}&=2.536\pm0.178,
 & B_{\rm NS,RSD}^{\rm II}&=12.63.
 \label{rsdbayesfactordefinition}
\end{align}
As in the no-RSD analysis, Convention~II is obtained by conditioning on the
remaining regular physical support:
\begin{align}
 \Delta\ln B_{\rm NS,RSD}^{\rm II}
 &=\Delta\ln B_{\rm NS,RSD}^{\rm I}-\ln f_{\rm root}^{\rm RSD}
 \nonumber \\
 &=2.53630\pm0.17786.
\end{align}
The additional binomial uncertainty,
$\sigma(\ln f_{\rm root}^{\rm RSD})=0.00535$, is negligible compared with
the nested-sampling uncertainty.  Thus, both prior conventions give
$\Delta\ln B_{\rm NS,RSD}>0$.

A separate matched calculation with $n_{\rm live}=400$ provides a numerical
check, giving
$\ln Z_{\rm SVT,RSD}^{\rm I}=-851.274\pm0.156$ and
$\ln Z_{\Lambda{\rm CDM,RSD}}^{\rm I}=-852.954\pm0.142$, and hence
$\Delta\ln B_{\rm NS,RSD}^{\rm I}=1.680\pm0.211$.  Its difference of $0.033$
from Eq.~(\ref{rsdbayesfactordefinition}) is far below the combined
numerical uncertainty.

\begin{figure*}[ht]
 \centering
 \includegraphics[width=0.82\textwidth]{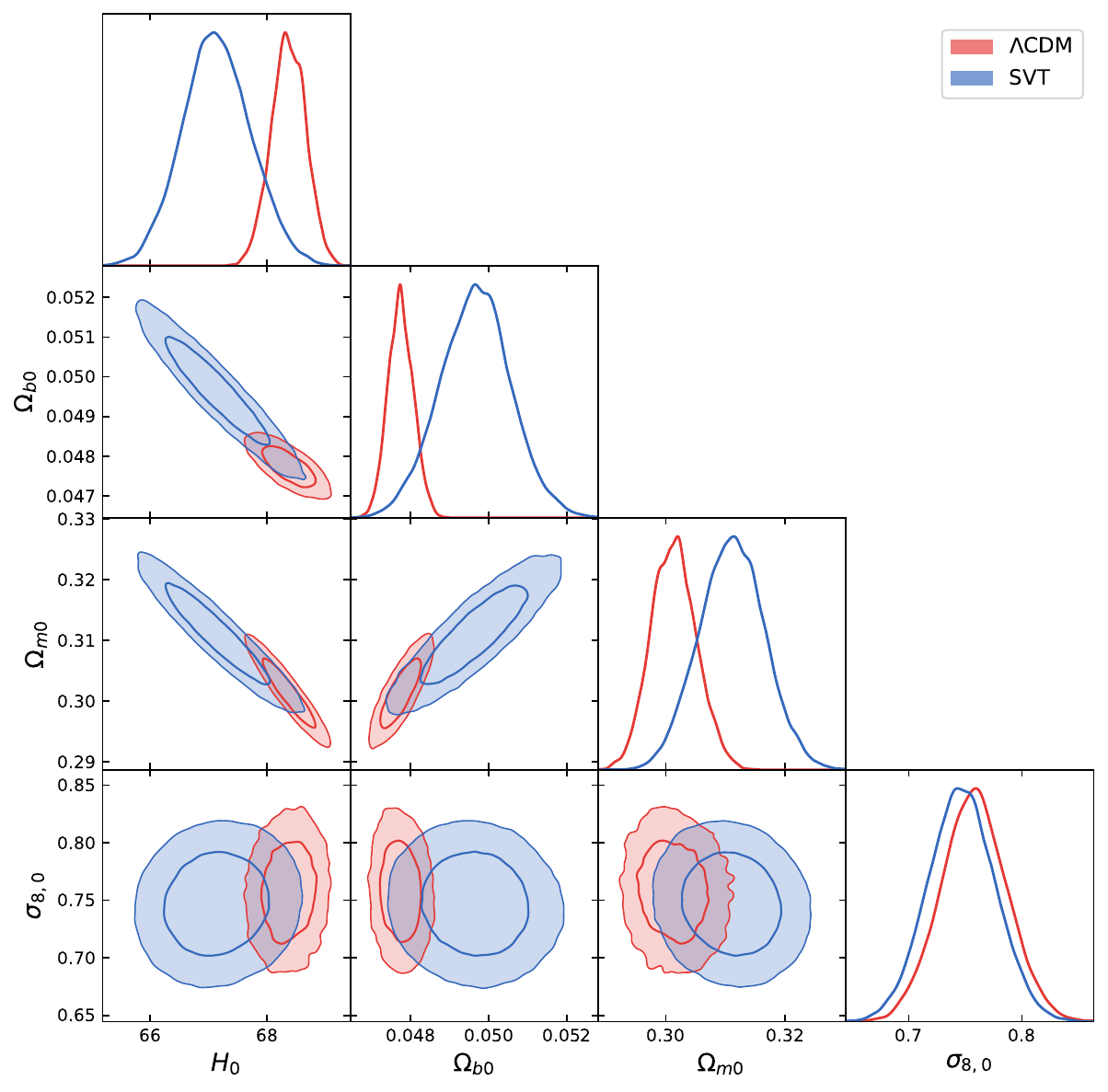}
 \caption{Matched posterior constraints on the four parameters common to SVT and flat
$\Lambda$CDM from DES-SN5YR+DESI DR2 BAO+compressed CMB+Gold-2017 RSD.
Red curves and contours correspond to $\Lambda$CDM and blue ones to SVT;
the inner and outer contours enclose the $68\%$ and $95\%$ credible regions,
respectively.  Relative to $\Lambda$CDM, the SVT posterior is shifted toward
smaller $H_0$ and larger $\Omega_{b0}$ and $\Omega_{m0}$, while the
$\sigma_{8,0}$ marginals overlap substantially.  These shifts are
parameter-estimation results and should not be interpreted as evidence for
one model over the other; the Bayesian model comparison is quantified
separately by Eq.~(\ref{rsdbayesfactordefinition}).  The contours are
obtained from the matched converged MCMC chains and are independent of the
evidence-normalization convention.}
 \label{fig:svtlcdmtrianglersd}
\end{figure*}

Figure~\ref{fig:svtlcdmtrianglersd} directly compares the four parameters
common to the SVT and $\Lambda$CDM RSD analyses. 
At 68\% credibility, the marginal intervals read
\begin{equation}
 \begin{array}{|c|c|c|}
 \hline
 &\Lambda{\rm CDM}&{\rm SVT}\\ \hline
 H_0&68.36^{+0.29}_{-0.29}&67.13^{+0.60}_{-0.56}\\
 \Omega_{b0}&0.04774^{+0.00034}_{-0.00032}&0.04966^{+0.00088}_{-0.00091}\\
 \Omega_{m0}&0.3013^{+0.0038}_{-0.0037}&0.3111^{+0.0053}_{-0.0055}\\
 \sigma_{8,0}&0.758^{+0.029}_{-0.029}&0.747^{+0.029}_{-0.029}\\
 \hline
 \end{array}
 \label{commonconstraintsrsd}
\end{equation}
Relative to flat $\Lambda$CDM, the SVT geometrical posterior is shifted
toward smaller $H_0$ and larger $\Omega_{b0}$ and $\Omega_{m0}$, as already
suggested by the no-RSD background fit.  In contrast, the fluctuation-amplitude
constraints overlap substantially: RSD determines $\sigma_{8,0}$ in both
models and places the SVT central value only mildly below the $\Lambda$CDM
one.  These contour displacements are parameter-estimation results and should
not themselves be interpreted as evidence for one model over the other; the
Bayesian model comparison is quantified instead by the matched evidence in
Eq.~(\ref{rsdbayesfactordefinition}).

%%%%%%%%%%%%%%%%%%%%%%%%%%%%%%%%%%%
%%%%%%%%%%%%%%%%%%%%%%%%%%%%%%%%%%%%%%%

Taken together, the RSD analysis preserves the reconstructed crossing behavior
within the perturbatively admissible posterior and constrains
$\sigma_{8,0}$ to values around $0.75$.  
The matched model comparison yields
$\Delta\chi^2_{\rm min,RSD}=-11.616$, 
$\Delta{\rm AIC}=-1.616$, and
$\Delta{\rm BIC}=+26.034$.  
The corresponding matched Bayes factors are
$B_{\rm NS,RSD}^{\rm I}=5.19$ and
$B_{\rm NS,RSD}^{\rm II}=12.63$ under 
the two prior conventions.

%%%%%%%%%%%%%%%%%%%%%%%%%%%%%
\section{Conclusions}
\label{consec}
%%%%%%%%%%%%%%%%%%%%%%%%%%%%%

We have tested the SVT DE model of
Ref.~\cite{Tsujikawa:2026xqm} against 
current measurements of the background
expansion and the growth of large-scale structure.  
In this model, the vector sector can drive 
the DE equation of state into the regime
$w_{\rm DE}<-1$ at earlier times, 
while a canonical scalar field evolving
along its potential subsequently drives $w_{\rm DE}$ through $-1$ toward
$w_{\rm DE}>-1$ at low redshift.  This realizes the crossing without
introducing a scalar field with a negative-sign kinetic term.  The model also
contains no nonminimal scalar or vector coupling to curvature, and the
restriction to cubic vector interactions keeps the tensor propagation speed
luminal.

Our baseline likelihood combines DES-SN5YR, DESI DR2 BAO, and compressed-CMB
information.  The background evolution is obtained by forward shooting from
the radiation era, with the vector density reconstructed algebraically, and
we sample the physical combinations $(f_\chi,s_\chi,\lambda)$ together with
the standard background parameters.  We then extend the analysis by adding
the 18-point Gold-2017 RSD compilation and the perturbation-sector parameters
$(p,\log_{10}\nu_v,\sigma_{8,0})$.  In the RSD analysis, we additionally
impose the perturbative support conditions of Sec.~\ref{perturbationtheory},
including positivity of the longitudinal-vector 
sound speed along the cosmological history.  
Converged four-chain MCMC runs are used for parameter
constraints and reconstructed histories, while matched PolyChord runs provide the Bayesian evidences.

For the no-RSD analysis, the posterior mean of the present DE equation of state is
$w_{{\rm DE},0}=-0.916457$, with a central 68\% interval
$[-0.943556,-0.884263]$.  The global best-fit history crosses
$w_{\rm DE}=-1$ near $z\simeq0.9$.  More directly, the weighted-history
estimator assigns
$P_{\rm cross}^{\rm no\mbox{-}RSD}=0.791276$ of the posterior weight to
regular histories satisfying $w_{\rm DE}(0)>-1$ and
$w_{\rm DE}(z)<-1$ at some $0<z\leq4$.  The best stored SVT point improves
the matched flat-$\Lambda$CDM fit by
$\Delta\chi^2_{\rm min}=-12.104$, with most of the improvement coming from
the supernova likelihood.  
The corresponding information-criterion
differences are $\Delta{\rm AIC}=-6.104$ and
$\Delta{\rm BIC}=+10.457$.

The RSD analysis leaves the reconstructed background evolution qualitatively
similar while adding direct information on structure growth.  In particular,
the SVT posterior gives
$\sigma_{8,0}=0.747^{+0.029}_{-0.029}$ at 68\% credibility.  The secondary
small-$f_\chi$ peak present in the no-RSD posterior disappears once the RSD
parameter space and perturbative support conditions are imposed.  The
directional crossing probability becomes
$P_{\rm cross}^{\rm RSD}=0.932745$.  This increase relative to the no-RSD
value should not be attributed to the RSD likelihood alone, because the RSD
analysis simultaneously imposes theoretical support conditions that are
absent from the six-dimensional background analysis.  The matched RSD
comparison gives
$\Delta\chi^2_{\rm min,RSD}=-11.616$,
$\Delta{\rm AIC}=-1.616$, and
$\Delta{\rm BIC}=+26.034$.  Thus the improvement in the best fit nearly
compensates for the AIC penalty associated with the five additional SVT
parameters, whereas the conventional BIC penalty is substantially larger.

For the Bayesian comparison, we report results under two explicit prior conventions.  
Under Convention~I, in which the rectangular sampled prior
retains the prior-volume penalty associated with points that fail to yield
regular shooting solutions, the no-RSD analysis gives
$\Delta\ln B_{\rm NS}^{\rm I}=0.720\pm0.163$
($B_{\rm NS}^{\rm I}=2.05$), while the RSD analysis gives
$\Delta\ln B_{\rm NS,RSD}^{\rm I}=1.647\pm0.178$
($B_{\rm NS,RSD}^{\rm I}=5.19$).  Under Convention~II, in which the SVT
prior is instead normalized on the remaining regular physical support, the
corresponding results are
$\Delta\ln B_{\rm NS}^{\rm II}=1.809\pm0.163$
($B_{\rm NS}^{\rm II}=6.11$) and
$\Delta\ln B_{\rm NS,RSD}^{\rm II}=2.536\pm0.178$
($B_{\rm NS,RSD}^{\rm II}=12.63$).  Dedicated prior-only calculations give
$f_{\rm root}^{\rm no\mbox{-}RSD}=0.33642\pm0.00211$ and
$f_{\rm root}^{\rm RSD}=0.41102\pm0.00220$, providing independent checks of
the prior-volume accounting.
The two conventions share the same posterior on the regular branch and differ
only in the normalization of its prior volume.  This normalization changes
the SVT Bayesian evidence, and hence the corresponding Bayes factor, but
leaves the posterior constraints, reconstructed $w_{\rm DE}(z)$ histories,
best-fit $\chi^2$, AIC, and BIC unchanged.

Taken together, the current background and growth data are compatible with
the SVT mechanism for crossing $w_{\rm DE}=-1$, 
with a substantial fraction
of the posterior weight assigned to the directional crossing histories defined in this work.  The matched PolyChord model comparison gives a Bayes
factor larger than unity for SVT relative to flat $\Lambda$CDM under both
prior conventions.  The Bayes factor is larger in the RSD analysis, although
that analysis also imposes the additional perturbative support conditions
discussed above.  At the same time, the contrasting indications from AIC,
BIC, and the directly computed Bayesian evidence, together with the
prior-convention dependence of the latter, highlight the importance of
specifying the prior measure explicitly in this theory-restricted parameter
space, where the $\Lambda$CDM limit is a nonregular boundary.

A full Boltzmann treatment of the CMB is an important next step, despite the
near-$\Lambda$CDM early-time behavior that motivates the present compressed-CMB analysis.  
Future higher-precision growth measurements,
including improved DESI RSD data and Euclid
galaxy-clustering and weak-lensing observations, will provide powerful tests of the SVT perturbation sector.  
In particular, the combination of clustering
and lensing can probe the effective gravitational coupling and the relation between the two metric potentials, thereby testing whether the crossing histories and growth-sector signatures found here persist with substantially
improved precision.

%%%%%%%%%%%%%%%%%%%%%%%%%%%%
\section*{Acknowledgements}
%%%%%%%%%%%%%%%%%%%%%%%%%%%%

S.T. acknowledges support from JSPS KAKENHI Grant Nos.~26K07090 and 26H00847 and from the Waseda University Special Research Projects (No.~2026C-486). S.T. also thanks the members of Tongji University for their warm hospitality during his visit, when part of this work was carried out. Y.Z. and Z.Z. are supported by the Fundamental Research Funds for the Central Universities, and by the National Natural Science Foundation of China (NSFC) Grants Nos. 12475060 and W2611007, Project 24ZR1472400 sponsored by Natural Science Foundation of Shanghai, and Shanghai Pujiang Program 24PJA134

\bibliographystyle{mybibstyle}
\bibliography{SVTconstraint_refs}

@article{Babichev:2011iz,
    author = "Babichev, Eugeny and Deffayet, Cedric and Esposito-Farese, Gilles",
    title = "{Constraints on Shift-Symmetric Scalar-Tensor Theories with a Vainshtein Mechanism from Bounds on the Time Variation of G}",
    eprint = "1107.1569",
    archivePrefix = "arXiv",
    primaryClass = "gr-qc",
    doi = "10.1103/PhysRevLett.107.251102",
    journal = "Phys. Rev. Lett.",
    volume = "107",
    pages = "251102",
    year = "2011"
}

@article{Kimura:2011dc,
    author = "Kimura, Rampei and Kobayashi, Tsutomu and Yamamoto, Kazuhiro",
    title = "{Vainshtein screening in a cosmological background in the most general second-order scalar-tensor theory}",
    eprint = "1111.6749",
    archivePrefix = "arXiv",
    primaryClass = "astro-ph.CO",
    doi = "10.1103/PhysRevD.85.024023",
    journal = "Phys. Rev. D",
    volume = "85",
    pages = "024023",
    year = "2012"
}

@article{Hofmann:2018myc,
    author = {Hofmann, F. and M{\"u}ller, J.},
    title = "{Relativistic tests with lunar laser ranging}",
    doi = "10.1088/1361-6382/aa8f7a",
    journal = "Class. Quant. Grav.",
    volume = "35",
    number = "3",
    pages = "035015",
    year = "2018"
}

@article{Tsujikawa:2019pih,
    author = "Tsujikawa, Shinji",
    title = "{Lunar Laser Ranging constraints on nonminimally coupled 
    dark energy and standard sirens}",
    eprint = "1903.07092",
    archivePrefix = "arXiv",
    primaryClass = "gr-qc",
    doi = "10.1103/PhysRevD.100.043510",
    journal = "Phys. Rev. D",
    volume = "100",
    number = "4",
    pages = "043510",
    year = "2019"
}

@article{Akaike:1974,
    author = "Akaike, Hirotugu",
    title = "{A new look at the statistical model identification}",
    journal = "IEEE Trans. Automatic Control",
    volume = "19",
    number = "6",
    pages = "716--723",
    year = "1974",
    doi = "10.1109/TAC.1974.1100705"
}

@article{Roy:2026icy,
    author = "Roy, Nandan and Sahoo, Prasanta",
    title = "{CosmoDS: A Python toolkit for constraining cosmological models via dynamical systems analysis with Cobaya}",
    eprint = "2603.14740",
    archivePrefix = "arXiv",
    primaryClass = "astro-ph.CO",
    month = "3",
    year = "2026"
}

@article{Schwarz:1978,
    author = "Schwarz, Gideon",
    title = "{Estimating the Dimension of a Model}",
    journal = "Annals Statist.",
    volume = "6",
    number = "2",
    pages = "461--464",
    year = "1978",
    doi = "10.1214/aos/1176344136"
}

@article{DeFelice:2010as,
    author = "De Felice, Antonio and Kase, Ryotaro and Tsujikawa, Shinji",
    title = "{Matter perturbations in Galileon cosmology}",
    eprint = "1011.6132",
    archivePrefix = "arXiv",
    primaryClass = "astro-ph.CO",
    doi = "10.1103/PhysRevD.83.043515",
    journal = "Phys. Rev. D",
    volume = "83",
    pages = "043515",
    year = "2011"
}

@article{Kaiser:1987qv,
    author = "Kaiser, N.",
    title = "{Clustering in real space and in redshift space}",
    doi = "10.1093/mnras/227.1.1",
    journal = "Mon. Not. Roy. Astron. Soc.",
    volume = "227",
    pages = "1--27",
    year = "1987"
}

@article{Song:2008qt,
    author = "Song, Yong-Seon and Percival, Will J.",
    title = "{Reconstructing the history of structure formation using Redshift Distortions}",
    eprint = "0807.0810",
    archivePrefix = "arXiv",
    primaryClass = "astro-ph",
    doi = "10.1088/1475-7516/2009/10/004",
    journal = "JCAP",
    volume = "10",
    pages = "004",
    year = "2009"
}

@article{Percival:2008sh,
    author = "Percival, Will J and White, Martin",
    title = "{Testing cosmological structure formation using redshift-space distortions}",
    eprint = "0808.0003",
    archivePrefix = "arXiv",
    primaryClass = "astro-ph",
    doi = "10.1111/j.1365-2966.2008.14211.x",
    journal = "Mon. Not. Roy. Astron. Soc.",
    volume = "393",
    pages = "297",
    year = "2009"
}

@article{Nakamura:2018oyy,
    author = "Nakamura, Shintaro and De Felice, Antonio and Kase, Ryotaro and Tsujikawa, Shinji",
    title = "{Constraints on massive vector dark energy models from integrated Sachs-Wolfe-galaxy cross-correlations}",
    eprint = "1811.07541",
    archivePrefix = "arXiv",
    primaryClass = "astro-ph.CO",
    doi = "10.1103/PhysRevD.99.063533",
    journal = "Phys. Rev. D",
    volume = "99",
    number = "6",
    pages = "063533",
    year = "2019"
}

@article{DeFelice:2020sdq,
    author = "De Felice, Antonio and Geng, Chao-Qiang and Pookkillath, Masroor C. and Yin, Lu",
    title = "{Reducing the $H_{0}$ tension with generalized Proca theory}",
    eprint = "2002.06782",
    archivePrefix = "arXiv",
    primaryClass = "astro-ph.CO",
    reportNumber = "YITP-20-65",
    doi = "10.1088/1475-7516/2020/08/038",
    journal = "JCAP",
    volume = "08",
    pages = "038",
    year = "2020"
}

@article{Heisenberg:2020xak,
    author = "Heisenberg, Lavinia and Villarrubia-Rojo, Hector",
    title = "{Proca in the sky}",
    eprint = "2010.00513",
    archivePrefix = "arXiv",
    primaryClass = "astro-ph.CO",
    doi = "10.1088/1475-7516/2021/03/032",
    journal = "JCAP",
    volume = "03",
    pages = "032",
    year = "2021"
}

@article{Silva:2009km,
    author = "Silva, Fabio P. and Koyama, Kazuya",
    title = "{Self-Accelerating Universe in Galileon Cosmology}",
    eprint = "0909.4538",
    archivePrefix = "arXiv",
    primaryClass = "astro-ph.CO",
    doi = "10.1103/PhysRevD.80.121301",
    journal = "Phys. Rev. D",
    volume = "80",
    pages = "121301",
    year = "2009"
}

@article{Deffayet:2010qz,
    author = "Deffayet, Cedric and Pujolas, Oriol and Sawicki, Ignacy and Vikman, Alexander",
    title = "{Imperfect Dark Energy from Kinetic Gravity Braiding}",
    eprint = "1008.0048",
    archivePrefix = "arXiv",
    primaryClass = "hep-th",
    reportNumber = "CERN-PH-TH-2010-166",
    doi = "10.1088/1475-7516/2010/10/026",
    journal = "JCAP",
    volume = "10",
    pages = "026",
    year = "2010"
}

@article{Chiba:1999ka,
    author = "Chiba, Takeshi and Okabe, Takahiro and Yamaguchi, Masahide",
    title = "{Kinetically driven quintessence}",
    eprint = "astro-ph/9912463",
    archivePrefix = "arXiv",
    reportNumber = "UTAP-352",
    doi = "10.1103/PhysRevD.62.023511",
    journal = "Phys. Rev. D",
    volume = "62",
    pages = "023511",
    year = "2000"
}

@article{Armendariz-Picon:2000ulo,
    author = "Armendariz-Picon, C. and Mukhanov, Viatcheslav F. and Steinhardt, Paul J.",
    title = "{Essentials of k essence}",
    eprint = "astro-ph/0006373",
    archivePrefix = "arXiv",
    doi = "10.1103/PhysRevD.63.103510",
    journal = "Phys. Rev. D",
    volume = "63",
    pages = "103510",
    year = "2001"
}

@article{Guo:2004fq,
    author = "Guo, Zong-Kuan and Piao, Yun-Song and Zhang, Xin-Min and Zhang, Yuan-Zhong",
    title = "{Cosmological evolution of a quintom model of dark energy}",
    eprint = "astro-ph/0410654",
    archivePrefix = "arXiv",
    doi = "10.1016/j.physletb.2005.01.017",
    journal = "Phys. Lett. B",
    volume = "608",
    pages = "177--182",
    year = "2005"
}

@article{Kobayashi:2010cm,
    author = "Kobayashi, Tsutomu and Yamaguchi, Masahide and Yokoyama, Jun'ichi",
    title = "{G-inflation: Inflation driven by the Galileon field}",
    eprint = "1008.0603",
    archivePrefix = "arXiv",
    primaryClass = "hep-th",
    reportNumber = "RESCEU-18-10",
    doi = "10.1103/PhysRevLett.105.231302",
    journal = "Phys. Rev. Lett.",
    volume = "105",
    pages = "231302",
    year = "2010"
}

@article{Tsujikawa:2026xqm,
    author = "Tsujikawa, Shinji",
    title = "{Realizing the phantom-divide crossing with vector and scalar fields}",
    eprint = "2601.21274",
    archivePrefix = "arXiv",
    primaryClass = "astro-ph.CO",
    reportNumber = "WUCG-26-01",
    doi = "10.1088/1475-7516/2026/06/009",
    journal = "JCAP",
    volume = "06",
    pages = "009",
    month = "1",
    year = "2026"
}

@article{Heisenberg:2014rta,
    author = "Heisenberg, Lavinia",
    title = "{Generalization of the Proca Action}",
    eprint = "1402.7026",
    archivePrefix = "arXiv",
    primaryClass = "hep-th",
    doi = "10.1088/1475-7516/2014/05/015",
    journal = "JCAP",
    volume = "05",
    pages = "015",
    year = "2014"
}

@article{Tasinato:2014eka,
    author = "Tasinato, Gianmassimo",
    title = "{Cosmic Acceleration from Abelian Symmetry Breaking}",
    eprint = "1402.6450",
    archivePrefix = "arXiv",
    primaryClass = "hep-th",
    doi = "10.1007/JHEP04(2014)067",
    journal = "JHEP",
    volume = "04",
    pages = "067",
    year = "2014"
}

@article{Allys:2015sht,
    author = "Allys, Erwan and Peter, Patrick and Rodriguez, Yeinzon",
    title = "{Generalized Proca action for an Abelian vector field}",
    eprint = "1511.03101",
    archivePrefix = "arXiv",
    primaryClass = "hep-th",
    reportNumber = "PI-UAN-2015-589FT",
    doi = "10.1088/1475-7516/2016/02/004",
    journal = "JCAP",
    volume = "02",
    pages = "004",
    year = "2016"
}

@article{BeltranJimenez:2016rff,
    author = "Beltran Jimenez, Jose and Heisenberg, Lavinia",
    title = "{Derivative self-interactions for a massive vector field}",
    eprint = "1602.03410",
    archivePrefix = "arXiv",
    primaryClass = "hep-th",
    doi = "10.1016/j.physletb.2016.04.017",
    journal = "Phys. Lett. B",
    volume = "757",
    pages = "405--411",
    year = "2016"
}

@article{Allys:2016jaq,
    author = "Allys, Erwan and Beltran Almeida, Juan P. and Peter, Patrick and Rodr\'\i{}guez, Yeinzon",
    title = "{On the 4D generalized Proca action for an Abelian vector field}",
    eprint = "1605.08355",
    archivePrefix = "arXiv",
    primaryClass = "hep-th",
    reportNumber = "PI-UAN-2016-595FT",
    doi = "10.1088/1475-7516/2016/09/026",
    journal = "JCAP",
    volume = "09",
    pages = "026",
    year = "2016"
}

@article{DeFelice:2016yws,
    author = "De Felice, Antonio and Heisenberg, Lavinia and Kase, 
    Ryotaro and Mukohyama, Shinji and Tsujikawa, Shinji and Zhang, Ying-li",
    title = "{Cosmology in generalized Proca theories}",
    eprint = "1603.05806",
    archivePrefix = "arXiv",
    primaryClass = "gr-qc",
    reportNumber = "YITP-16-36, IPMU16-0032",
    doi = "10.1088/1475-7516/2016/06/048",
    journal = "JCAP",
    volume = "06",
    pages = "048",
    year = "2016"
}

@article{DeFelice:2016uil,
    author = "De Felice, Antonio and Heisenberg, Lavinia and Kase, Ryotaro and Mukohyama, Shinji and Tsujikawa, Shinji and Zhang, Ying-li",
    title = "{Effective gravitational couplings for cosmological perturbations
     in generalized Proca theories}",
    eprint = "1605.05066",
    archivePrefix = "arXiv",
    primaryClass = "gr-qc",
    doi = "10.1103/PhysRevD.94.044024",
    journal = "Phys. Rev. D",
    volume = "94",
    number = "4",
    pages = "044024",
    year = "2016"
}

@article{deFelice:2017paw,
    author = "De Felice, Antonio and Heisenberg, Lavinia and Tsujikawa, Shinji",
    title = "{Observational constraints on generalized Proca theories}",
    eprint = "1703.09573",
    archivePrefix = "arXiv",
    primaryClass = "astro-ph.CO",
    doi = "10.1103/PhysRevD.95.123540",
    journal = "Phys. Rev. D",
    volume = "95",
    number = "12",
    pages = "123540",
    year = "2017"
}

@article{Tsujikawa:2025wca,
    author = "Tsujikawa, Shinji",
    title = "{Crossing the phantom divide in scalar-tensor and vector-tensor theories}",
    eprint = "2508.17231",
    archivePrefix = "arXiv",
    primaryClass = "astro-ph.CO",
    reportNumber = "WUCG-25-09",
    doi = "10.1103/y858-4swl",
    journal = "Phys. Rev. D",
    volume = "113",
    number = "4",
    pages = "L041301",
    year = "2026"
}

@article{Torrado:2020dgo,
    author = "Torrado, Jesus and Lewis, Antony",
    title = "{Cobaya: Code for Bayesian Analysis of hierarchical physical models}",
    eprint = "2005.05290",
    archivePrefix = "arXiv",
    primaryClass = "astro-ph.IM",
    reportNumber = "TTK-20-15",
    doi = "10.1088/1475-7516/2021/05/057",
    journal = "JCAP",
    volume = "05",
    pages = "057",
    year = "2021"
}

@article{Lewis:2002ah,
 author = {Lewis, Antony and Bridle, Sarah},
 title = {Cosmological parameters from CMB and other data: A Monte Carlo approach},
 journal = {Phys. Rev. D}, volume = {66}, pages = {103511}, year = {2002},
 doi = {10.1103/PhysRevD.66.103511},
 eprint = {astro-ph/0205436}, archivePrefix = {arXiv}
}

@article{Lewis:2013hha,
 author = {Lewis, Antony},
 title = {Efficient sampling of fast and slow cosmological parameters},
 journal = {Phys. Rev. D}, volume = {87}, pages = {103529}, year = {2013},
 doi = {10.1103/PhysRevD.87.103529},
 eprint = {1304.4473}, archivePrefix = {arXiv}, primaryClass = {astro-ph.CO}
}

@article{Lewis:2019xzd,
    author = "Lewis, Antony",
    title = "{GetDist: a Python package for analysing Monte Carlo samples}",
    eprint = "1910.13970",
    archivePrefix = "arXiv",
    primaryClass = "astro-ph.IM",
    month = "10",
    year = "2019"
}

@article{DES:2024jxu,
    author = "Abbott, T. M. C. and others",
    collaboration = "DES",
    title = "{The Dark Energy Survey: Cosmology Results with \ensuremath{\sim}1500 New High-redshift Type Ia Supernovae Using the Full 5 yr Data Set}",
    eprint = "2401.02929",
    archivePrefix = "arXiv",
    primaryClass = "astro-ph.CO",
    reportNumber = "FERMILAB-PUB-23-0821-PPD, DES-2023-805",
    doi = "10.3847/2041-8213/ad6f9f",
    journal = "Astrophys. J. Lett.",
    volume = "973",
    number = "1",
    pages = "L14",
    year = "2024"
}

@article{DESI:2025zgx,
    author = "Abdul Karim, M. and others",
    collaboration = "DESI",
    title = "{DESI DR2 Results II: Measurements of 
    Baryon Acoustic Oscillations and Cosmological Constraints}",
    eprint = "2503.14738",
    archivePrefix = "arXiv",
    primaryClass = "astro-ph.CO",
    reportNumber = "FERMILAB-PUB-25-0169-PPD",
    month = "3",
    year = "2025"
}

@article{Lemos:2023xhs,
    author = "Lemos, Pablo and Lewis, Antony",
    title = "{CMB constraints on the early Universe independent of late-time cosmology}",
    eprint = "2302.12911",
    archivePrefix = "arXiv",
    primaryClass = "astro-ph.CO",
    doi = "10.1103/PhysRevD.107.103505",
    journal = "Phys. Rev. D",
    volume = "107",
    number = "10",
    pages = "103505",
    year = "2023"
}

@article{Lewis:1999bs,
    author = "Lewis, Antony and Challinor, Anthony and Lasenby, Anthony",
    title = "{Efficient computation of cosmic microwave background anisotropies in closed Friedmann-Robertson-Walker models}",
    eprint = "astro-ph/9911177",
    archivePrefix = "arXiv",
    doi = "10.1086/309179",
    journal = "Astrophys. J.",
    volume = "538",
    pages = "473--476",
    year = "2000"
}

@article{Nesseris:2017vor,
    author = "Nesseris, Savvas and Pantazis, George and Perivolaropoulos, Leandros",
    title = "{Tension and constraints on modified gravity parametrizations of $G_{\textrm{eff}}(z)$ from growth rate and Planck data}",
    eprint = "1703.10538",
    archivePrefix = "arXiv",
    primaryClass = "astro-ph.CO",
    reportNumber = "IFT-UAM-CSIC-17-031",
    doi = "10.1103/PhysRevD.96.023542",
    journal = "Phys. Rev. D",
    volume = "96",
    number = "2",
    pages = "023542",
    year = "2017"
}

@article{Peebles:2002gy,
    author = "Peebles, P. J. E. and Ratra, Bharat",
    editor = "Hsu, Jong-Ping and Fine, D.",
    title = "{The Cosmological Constant and Dark Energy}",
    eprint = "astro-ph/0207347",
    archivePrefix = "arXiv",
    reportNumber = "KSUPT-02-3",
    doi = "10.1103/RevModPhys.75.559",
    journal = "Rev. Mod. Phys.",
    volume = "75",
    pages = "559--606",
    year = "2003"
}

@article{SDSS:2005xqv,
    author = "Eisenstein, Daniel J. and others",
    collaboration = "SDSS",
    title = "{Detection of the Baryon Acoustic Peak in the Large-Scale Correlation Function of SDSS Luminous Red Galaxies}",
    eprint = "astro-ph/0501171",
    archivePrefix = "arXiv",
    reportNumber = "FERMILAB-PUB-05-057-A-CD",
    doi = "10.1086/466512",
    journal = "Astrophys. J.",
    volume = "633",
    pages = "560--574",
    year = "2005"
}

@article{Heisenberg:2018mxx,
    author = "Heisenberg, Lavinia and Kase, Ryotaro and Tsujikawa, Shinji",
    title = "{Cosmology in scalar-vector-tensor theories}",
    eprint = "1805.01066",
    archivePrefix = "arXiv",
    primaryClass = "gr-qc",
    doi = "10.1103/PhysRevD.98.024038",
    journal = "Phys. Rev. D",
    volume = "98",
    number = "2",
    pages = "024038",
    year = "2018"
}

@article{Linder:2002et,
    author = "Linder, Eric V.",
    title = "{Exploring the expansion history of the universe}",
    eprint = "astro-ph/0208512",
    archivePrefix = "arXiv",
    doi = "10.1103/PhysRevLett.90.091301",
    journal = "Phys. Rev. Lett.",
    volume = "90",
    pages = "091301",
    year = "2003"
}

@article{DeFelice:2010nf,
    author = "De Felice, Antonio and Tsujikawa, Shinji",
    title = "{Generalized Galileon cosmology}",
    eprint = "1008.4236",
    archivePrefix = "arXiv",
    primaryClass = "hep-th",
    doi = "10.1103/PhysRevD.84.124029",
    journal = "Phys. Rev. D",
    volume = "84",
    pages = "124029",
    year = "2011"
}

@article{Aoki:2024ktc,
    author = "Aoki, Katsuki and Gorji, Mohammad Ali and Hiramatsu, Takashi and Mukohyama, Shinji and Pookkillath, Masroor C. and Takahashi, Kazufumi",
    title = "{CMB spectrum in unified EFT of dark energy: scalar-tensor and vector-tensor theories}",
    eprint = "2405.04265",
    archivePrefix = "arXiv",
    primaryClass = "astro-ph.CO",
    reportNumber = "YITP-24-56, RUP-24-8, IPMU24-0016",
    doi = "10.1088/1475-7516/2024/07/056",
    journal = "JCAP",
    volume = "07",
    pages = "056",
    year = "2024"
}

@article{Aoki:2021wew,
    author = "Aoki, Katsuki and Gorji, Mohammad Ali and Mukohyama, Shinji and Takahashi, Kazufumi",
    title = "{The effective field theory of vector-tensor theories}",
    eprint = "2111.08119",
    archivePrefix = "arXiv",
    primaryClass = "hep-th",
    reportNumber = "YITP-21-132, IPMU21-0079",
    doi = "10.1088/1475-7516/2022/01/059",
    journal = "JCAP",
    volume = "01",
    number = "01",
    pages = "059",
    year = "2022"
}

@article{Heisenberg:2018acv,
    author = "Heisenberg, Lavinia",
    title = "{Scalar-Vector-Tensor Gravity Theories}",
    eprint = "1801.01523",
    archivePrefix = "arXiv",
    primaryClass = "gr-qc",
    doi = "10.1088/1475-7516/2018/10/054",
    journal = "JCAP",
    volume = "10",
    pages = "054",
    year = "2018"
}

@article{Armendariz-Picon:2000nqq,
    author = "Armendariz-Picon, C. and Mukhanov, Viatcheslav F. and Steinhardt, Paul J.",
    title = "{A Dynamical solution to the problem of a small cosmological constant and late time cosmic acceleration}",
    eprint = "astro-ph/0004134",
    archivePrefix = "arXiv",
    doi = "10.1103/PhysRevLett.85.4438",
    journal = "Phys. Rev. Lett.",
    volume = "85",
    pages = "4438--4441",
    year = "2000"
}

@article{SupernovaSearchTeam:1998fmf,
    author = "Riess, Adam G. and others",
    collaboration = "Supernova Search Team",
    title = "{Observational evidence from supernovae for an accelerating universe and a cosmological constant}",
    eprint = "astro-ph/9805201",
    archivePrefix = "arXiv",
    doi = "10.1086/300499",
    journal = "Astron. J.",
    volume = "116",
    pages = "1009--1038",
    year = "1998"
}

@article{DESI:2024mwx,
    author = "Adame, A. G. and others",
    collaboration = "DESI",
    title = "{DESI 2024 VI: cosmological constraints from the measurements of baryon acoustic oscillations}",
    eprint = "2404.03002",
    archivePrefix = "arXiv",
    primaryClass = "astro-ph.CO",
    reportNumber = "FERMILAB-PUB-24-0154-PPD",
    doi = "10.1088/1475-7516/2025/02/021",
    journal = "JCAP",
    volume = "02",
    pages = "021",
    year = "2025"
}

@article{DESI:2025fii,
    author = "Lodha, K. and others",
    collaboration = "DESI",
    title = "{Extended dark energy analysis using DESI DR2 BAO measurements}",
    eprint = "2503.14743",
    archivePrefix = "arXiv",
    primaryClass = "astro-ph.CO",
    reportNumber = "FERMILAB-PUB-25-0164-PPD",
    doi = "10.1103/w4c6-1r5j",
    journal = "Phys. Rev. D",
    volume = "112",
    number = "8",
    pages = "083511",
    year = "2025"
}

@article{Deffayet:2009wt,
    author = "Deffayet, C. and Esposito-Farese, Gilles and Vikman, A.",
    title = "{Covariant Galileon}",
    eprint = "0901.1314",
    archivePrefix = "arXiv",
    primaryClass = "hep-th",
    doi = "10.1103/PhysRevD.79.084003",
    journal = "Phys. Rev. D",
    volume = "79",
    pages = "084003",
    year = "2009"
}

@article{WMAP:2003elm,
    author = "Spergel, D. N. and others",
    collaboration = "WMAP",
    title = "{First year Wilkinson Microwave Anisotropy Probe (WMAP) observations: Determination of cosmological parameters}",
    eprint = "astro-ph/0302209",
    archivePrefix = "arXiv",
    doi = "10.1086/377226",
    journal = "Astrophys. J. Suppl.",
    volume = "148",
    pages = "175--194",
    year = "2003"
}

@article{Heisenberg:2018wye,
    author = "Heisenberg, Lavinia and Kase, Ryotaro and Tsujikawa, Shinji",
    title = "{Gauge-ready formulation of cosmological perturbations in scalar-vector-tensor theories}",
    eprint = "1807.07202",
    archivePrefix = "arXiv",
    primaryClass = "gr-qc",
    doi = "10.1103/PhysRevD.98.123504",
    journal = "Phys. Rev. D",
    volume = "98",
    number = "12",
    pages = "123504",
    year = "2018"
}

@article{Aoki:2025bmj,
    author = "Aoki, Katsuki and Beltr{\'a}n Jim{\'e}nez, Jose and Pookkillath, Masroor C. and Tsujikawa, Shinji",
    title = "{Effective field theory of coupled dark energy and dark matter}",
    eprint = "2504.17293",
    archivePrefix = "arXiv",
    primaryClass = "astro-ph.CO",
    reportNumber = "YITP-25-56, WUCG-25-04",
    doi = "10.1103/5wsb-1vk6",
    journal = "Phys. Rev. D",
    volume = "113",
    number = "4",
    pages = "044053",
    year = "2026"
}

@article{Nicolis:2008in,
    author = "Nicolis, Alberto and Rattazzi, Riccardo and Trincherini, Enrico",
    title = "{The Galileon as a local modification of gravity}",
    eprint = "0811.2197",
    archivePrefix = "arXiv",
    primaryClass = "hep-th",
    doi = "10.1103/PhysRevD.79.064036",
    journal = "Phys. Rev. D",
    volume = "79",
    pages = "064036",
    year = "2009"
}

@article{Wetterich:1987fm,
    author = "Wetterich, C.",
    title = "{Cosmology and the Fate of Dilatation Symmetry}",
    eprint = "1711.03844",
    archivePrefix = "arXiv",
    primaryClass = "hep-th",
    reportNumber = "PRINT-87-0756, DESY-87-123",
    doi = "10.1016/0550-3213(88)90193-9",
    journal = "Nucl. Phys. B",
    volume = "302",
    pages = "668--696",
    year = "1988"
}

@article{Cline:2003gs,
    author = "Cline, James M. and Jeon, Sangyong and Moore, Guy D.",
    title = "{The Phantom menaced: Constraints on low-energy effective ghosts}",
    eprint = "hep-ph/0311312",
    archivePrefix = "arXiv",
    reportNumber = "MCGILL-03-25",
    doi = "10.1103/PhysRevD.70.043543",
    journal = "Phys. Rev. D",
    volume = "70",
    pages = "043543",
    year = "2004"
}

@article{DESI:2024aqx,
    author = "Calderon, R. and others",
    collaboration = "DESI",
    title = "{DESI 2024: reconstructing dark energy using crossing statistics with DESI DR1 BAO data}",
    eprint = "2405.04216",
    archivePrefix = "arXiv",
    primaryClass = "astro-ph.CO",
    doi = "10.1088/1475-7516/2024/10/048",
    journal = "JCAP",
    volume = "10",
    pages = "048",
    year = "2024"
}

@article{Copeland:2006wr,
    author = "Copeland, Edmund J. and Sami, M. and Tsujikawa, Shinji",
    title = "{Dynamics of dark energy}",
    eprint = "hep-th/0603057",
    archivePrefix = "arXiv",
    doi = "10.1142/S021827180600942X",
    journal = "Int. J. Mod. Phys. D",
    volume = "15",
    pages = "1753--1936",
    year = "2006"
}

@article{Nesseris:2010pc,
    author = "Nesseris, Savvas and De Felice, Antonio and Tsujikawa, Shinji",
    title = "{Observational constraints on Galileon cosmology}",
    eprint = "1010.0407",
    archivePrefix = "arXiv",
    primaryClass = "astro-ph.CO",
    doi = "10.1103/PhysRevD.82.124054",
    journal = "Phys. Rev. D",
    volume = "82",
    pages = "124054",
    year = "2010"
}

@article{Feng:2004ad,
    author = "Feng, Bo and Wang, Xiu-Lian and Zhang, Xin-Min",
    title = "{Dark energy constraints from the cosmic age and supernova}",
    eprint = "astro-ph/0404224",
    archivePrefix = "arXiv",
    doi = "10.1016/j.physletb.2004.12.071",
    journal = "Phys. Lett. B",
    volume = "607",
    pages = "35--41",
    year = "2005"
}

@article{Ratra:1987rm,
    author = "Ratra, Bharat and Peebles, P. J. E.",
    title = "{Cosmological Consequences of a Rolling Homogeneous Scalar Field}",
    reportNumber = "PUPT-1072",
    doi = "10.1103/PhysRevD.37.3406",
    journal = "Phys. Rev. D",
    volume = "37",
    pages = "3406",
    year = "1988"
}

@article{Simpson:2012ra,
    author = "Simpson, Fergus and others",
    title = "{CFHTLenS: Testing the Laws of Gravity with Tomographic Weak Lensing and Redshift Space Distortions}",
    eprint = "1212.3339",
    archivePrefix = "arXiv",
    primaryClass = "astro-ph.CO",
    doi = "10.1093/mnras/sts493",
    journal = "Mon. Not. Roy. Astron. Soc.",
    volume = "429",
    pages = "2249",
    year = "2013"
}

@article{Peirone:2019aua,
    author = "Peirone, Simone and Benevento, Giampaolo and Frusciante, Noemi and Tsujikawa, Shinji",
    title = "{Cosmological data favor Galileon ghost condensate over $\Lambda$CDM}",
    eprint = "1905.05166",
    archivePrefix = "arXiv",
    primaryClass = "astro-ph.CO",
    doi = "10.1103/PhysRevD.100.063540",
    journal = "Phys. Rev. D",
    volume = "100",
    number = "6",
    pages = "063540",
    year = "2019"
}

@article{DeFelice:2010pv,
    author = "De Felice, Antonio and Tsujikawa, Shinji",
    title = "{Cosmology of a covariant Galileon field}",
    eprint = "1007.2700",
    archivePrefix = "arXiv",
    primaryClass = "astro-ph.CO",
    doi = "10.1103/PhysRevLett.105.111301",
    journal = "Phys. Rev. Lett.",
    volume = "105",
    pages = "111301",
    year = "2010"
}

@article{Horndeski:1974wa,
    author = "Horndeski, Gregory Walter",
    title = "{Second-order scalar-tensor field equations in a four-dimensional space}",
    doi = "10.1007/BF01807638",
    journal = "Int. J. Theor. Phys.",
    volume = "10",
    pages = "363--384",
    year = "1974"
}

@article{Chevallier:2000qy,
    author = "Chevallier, Michel and Polarski, David",
    title = "{Accelerating universes with scaling dark matter}",
    eprint = "gr-qc/0009008",
    archivePrefix = "arXiv",
    doi = "10.1142/S0218271801000822",
    journal = "Int. J. Mod. Phys. D",
    volume = "10",
    pages = "213--224",
    year = "2001"
}

@article{Amendola:2007rr,
    author = "Amendola, Luca and Kunz, Martin and Sapone, Domenico",
    title = "{Measuring the dark side (with weak lensing)}",
    eprint = "0704.2421",
    archivePrefix = "arXiv",
    primaryClass = "astro-ph",
    doi = "10.1088/1475-7516/2008/04/013",
    journal = "JCAP",
    volume = "04",
    pages = "013",
    year = "2008"
}

@article{Kase:2018nwt,
    author = "Kase, Ryotaro and Tsujikawa, Shinji",
    title = "{Dark energy in scalar-vector-tensor theories}",
    eprint = "1805.11919",
    archivePrefix = "arXiv",
    primaryClass = "gr-qc",
    doi = "10.1088/1475-7516/2018/11/024",
    journal = "JCAP",
    volume = "11",
    pages = "024",
    year = "2018"
}

@article{Carroll:2003st,
    author = "Carroll, Sean M. and Hoffman, Mark and Trodden, Mark",
    title = "{Can the dark energy equation-of-state parameter $w$ be less than $-1$?}",
    eprint = "astro-ph/0301273",
    archivePrefix = "arXiv",
    reportNumber = "EFI-2003-01, SU-GP-03-1-1",
    doi = "10.1103/PhysRevD.68.023509",
    journal = "Phys. Rev. D",
    volume = "68",
    pages = "023509",
    year = "2003"
}

@article{Caldwell:1999ew,
    author = "Caldwell, R. R.",
    title = "{A Phantom menace?}",
    eprint = "astro-ph/9908168",
    archivePrefix = "arXiv",
    doi = "10.1016/S0370-2693(02)02589-3",
    journal = "Phys. Lett. B",
    volume = "545",
    pages = "23--29",
    year = "2002"
}

@article{SupernovaCosmologyProject:1998vns,
    author = "Perlmutter, S. and others",
    collaboration = "Supernova Cosmology Project",
    title = "{Measurements of $\Omega$ and $\Lambda$ from 42 High Redshift Supernovae}",
    eprint = "astro-ph/9812133",
    archivePrefix = "arXiv",
    reportNumber = "LBNL-41801, LBL-41801",
    doi = "10.1086/307221",
    journal = "Astrophys. J.",
    volume = "517",
    pages = "565--586",
    year = "1999"
}

@article{Handley:2015polychord,
    author = "Handley, W. J. and Hobson, M. P. and Lasenby, A. N.",
    title = "{PolyChord: nested sampling for cosmology}",
    eprint = "1502.01856",
    archivePrefix = "arXiv",
    primaryClass = "astro-ph.CO",
    doi = "10.1093/mnrasl/slv047",
    journal = "Mon. Not. Roy. Astron. Soc.",
    volume = "450",
    number = "1",
    pages = "L61--L65",
    year = "2015"
}

@article{SilvestriTrodden:2009,
    author = "Silvestri, Alessandra and Trodden, Mark",
    title = "{Approaches to Understanding Cosmic Acceleration}",
    eprint = "0904.0024",
    archivePrefix = "arXiv",
    primaryClass = "astro-ph.CO",
    doi = "10.1088/0034-4885/72/9/096901",
    journal = "Rept. Prog. Phys.",
    volume = "72",
    pages = "096901",
    year = "2009"
}

@article{CliftonEtAl:2012,
    author = "Clifton, Timothy and Ferreira, Pedro G. and Padilla, Antonio and Skordis, Constantinos",
    title = "{Modified Gravity and Cosmology}",
    eprint = "1106.2476",
    archivePrefix = "arXiv",
    primaryClass = "astro-ph.CO",
    doi = "10.1016/j.physrep.2012.01.001",
    journal = "Phys. Rept.",
    volume = "513",
    pages = "1--189",
    year = "2012"
}

@article{Tsujikawa:2013Quintessence,
    author = "Tsujikawa, Shinji",
    title = "{Quintessence: A Review}",
    eprint = "1304.1961",
    archivePrefix = "arXiv",
    primaryClass = "gr-qc",
    doi = "10.1088/0264-9381/30/21/214003",
    journal = "Class. Quant. Grav.",
    volume = "30",
    pages = "214003",
    year = "2013"
}

@article{JoyceEtAl:2015,
    author = "Joyce, Austin and Jain, Bhuvnesh and Khoury, Justin and Trodden, Mark",
    title = "{Beyond the Cosmological Standard Model}",
    eprint = "1407.0059",
    archivePrefix = "arXiv",
    primaryClass = "astro-ph.CO",
    doi = "10.1016/j.physrep.2014.12.002",
    journal = "Phys. Rept.",
    volume = "568",
    pages = "1--98",
    year = "2015"
}

@article{Koyama:2016,
    author = "Koyama, Kazuya",
    title = "{Cosmological Tests of Modified Gravity}",
    eprint = "1504.04623",
    archivePrefix = "arXiv",
    primaryClass = "astro-ph.CO",
    doi = "10.1088/0034-4885/79/4/046902",
    journal = "Rept. Prog. Phys.",
    volume = "79",
    pages = "046902",
    year = "2016"
}

@article{BullEtAl:2016,
    author = "Bull, Philip and others",
    title = "{Beyond {$\\Lambda$CDM}: Problems, solutions, and the road ahead}",
    eprint = "1512.05356",
    archivePrefix = "arXiv",
    primaryClass = "astro-ph.CO",
    doi = "10.1016/j.dark.2016.02.001",
    journal = "Phys. Dark Univ.",
    volume = "12",
    pages = "56--99",
    year = "2016"
}

@article{Heisenberg:2019Review,
    author = "Heisenberg, Lavinia",
    title = "{A systematic approach to generalisations of General Relativity and their cosmological implications}",
    eprint = "1807.01725",
    archivePrefix = "arXiv",
    primaryClass = "gr-qc",
    doi = "10.1016/j.physrep.2018.11.006",
    journal = "Phys. Rept.",
    volume = "796",
    pages = "1--113",
    year = "2019"
}

@article{KaseTsujikawa:2019Review,
    author = "Kase, Ryotaro and Tsujikawa, Shinji",
    title = "{Dark energy in Horndeski theories after GW170817: A review}",
    eprint = "1809.08735",
    archivePrefix = "arXiv",
    primaryClass = "gr-qc",
    doi = "10.1142/S0218271819420057",
    journal = "Int. J. Mod. Phys. D",
    volume = "28",
    number = "05",
    pages = "1942005",
    year = "2019"
}

@article{Fujii:1982,
    author = "Fujii, Yasunori",
    title = "{Origin of the Gravitational Constant and Particle Masses in a Scale-Invariant Scalar-Tensor Theory}",
    doi = "10.1103/PhysRevD.26.2580",
    journal = "Phys. Rev. D",
    volume = "26",
    pages = "2580",
    year = "1982"
}

@article{Chiba:1997,
    author = "Chiba, Takeshi and Sugiyama, Naoshi and Nakamura, Takashi",
    title = "{Cosmology with x-matter}",
    eprint = "astro-ph/9704199",
    archivePrefix = "arXiv",
    doi = "10.1093/mnras/289.2.L5",
    journal = "Mon. Not. Roy. Astron. Soc.",
    volume = "289",
    pages = "L5--L9",
    year = "1997"
}

@article{FerreiraJoyce:1997,
    author = "Ferreira, Pedro G. and Joyce, Michael",
    title = "{Structure Formation with a Self-Tuning Scalar Field}",
    eprint = "astro-ph/9707286",
    archivePrefix = "arXiv",
    doi = "10.1103/PhysRevLett.79.4740",
    journal = "Phys. Rev. Lett.",
    volume = "79",
    pages = "4740--4743",
    year = "1997"
}

@article{CaldwellDaveSteinhardt:1998,
    author = "Caldwell, R. R. and Dave, Rahul and Steinhardt, Paul J.",
    title = "{Cosmological Imprint of an Energy Component with General Equation of State}",
    eprint = "astro-ph/9708069",
    archivePrefix = "arXiv",
    doi = "10.1103/PhysRevLett.80.1582",
    journal = "Phys. Rev. Lett.",
    volume = "80",
    pages = "1582--1585",
    year = "1998"
}

@article{CopelandLiddleWands:1998,
    author = "Copeland, Edmund J. and Liddle, Andrew R. and Wands, David",
    title = "{Exponential Potentials and Cosmological Scaling Solutions}",
    eprint = "gr-qc/9711068",
    archivePrefix = "arXiv",
    doi = "10.1103/PhysRevD.57.4686",
    journal = "Phys. Rev. D",
    volume = "57",
    pages = "4686--4690",
    year = "1998"
}

@article{Ye:2024desi,
    author = "Ye, Gen and Martinelli, Matteo and Hu, Bin and Silvestri, Alessandra",
    title = "{Hints of Nonminimally Coupled Gravity in DESI 2024 Baryon Acoustic Oscillation Measurements}",
    eprint = "2407.15832",
    archivePrefix = "arXiv",
    primaryClass = "astro-ph.CO",
    doi = "10.1103/PhysRevLett.134.181002",
    journal = "Phys. Rev. Lett.",
    volume = "134",
    pages = "181002",
    year = "2025"
}

@article{Wolf:2024nmc,
    author = "Wolf, William J. and Ferreira, Pedro G. and Garc{\'i}a-Garc{\'i}a, Carlos",
    title = "{Matching current observational constraints with nonminimally coupled dark energy}",
    eprint = "2409.17019",
    archivePrefix = "arXiv",
    primaryClass = "astro-ph.CO",
    doi = "10.1103/PhysRevD.111.L041303",
    journal = "Phys. Rev. D",
    volume = "111",
    pages = "L041303",
    year = "2025"
}

@article{Ye:2024tg,
    author = "Ye, Gen",
    title = "{Bridge the Cosmological Tensions with Thawing Gravity}",
    eprint = "2411.11743",
    archivePrefix = "arXiv",
    primaryClass = "astro-ph.CO",
    year = "2024"
}

@article{Pan:2025desi,
    author = "Pan, Jiaming and Ye, Gen",
    title = "{Nonminimally coupled gravity constraints from DESI DR2 data}",
    eprint = "2503.19898",
    archivePrefix = "arXiv",
    primaryClass = "astro-ph.CO",
    doi = "10.1103/hqwq-m19h",
    journal = "Phys. Rev. D",
    volume = "113",
    pages = "L041304",
    year = "2026"
}

@article{Wolf:2025nmc,
    author = "Wolf, William J. and Garc{\'i}a-Garc{\'i}a, Carlos and Anton, Theodore and Ferreira, Pedro G.",
    title = "{Assessing Cosmological Evidence for Nonminimal Coupling}",
    eprint = "2504.07679",
    archivePrefix = "arXiv",
    primaryClass = "astro-ph.CO",
    doi = "10.1103/jysf-k72m",
    journal = "Phys. Rev. Lett.",
    volume = "135",
    pages = "081001",
    year = "2025"
}

@article{Wang:2025nmc,
    author = "Wang, Jia-Qi and Cai, Rong-Gen and Guo, Zong-Kuan and Wang, Shao-Jiang",
    title = "{Resolving the Planck-DESI tension by nonminimally coupled quintessence}",
    eprint = "2508.01759",
    archivePrefix = "arXiv",
    primaryClass = "astro-ph.CO",
    doi = "10.1103/r6cx-8ghz",
    journal = "Phys. Rev. D",
    volume = "113",
    pages = "083534",
    year = "2026"
}

@article{Adam:2025nmc,
    author = "Adam, Husam and Hertzberg, Mark P. and Jim{\'e}nez-Aguilar, Daniel and Khan, Iman",
    title = "{Comparing minimal and non-minimal quintessence models to 2025 DESI data}",
    eprint = "2509.13302",
    archivePrefix = "arXiv",
    primaryClass = "astro-ph.CO",
    doi = "10.1088/1475-7516/2026/04/052",
    journal = "JCAP",
    volume = "04",
    pages = "052",
    year = "2026"
}

@article{SanchezLopez:2025nmc,
    author = "S{\'a}nchez L{\'o}pez, Samuel and Karam, Alexandros and Hazra, Dhiraj Kumar",
    title = "{Non-Minimally Coupled Quintessence in Light of DESI}",
    eprint = "2510.14941",
    archivePrefix = "arXiv",
    primaryClass = "astro-ph.CO",
    year = "2025"
}

@article{Pookkillath:2026kinetic,
    author = "Pookkillath, Masroor C. and Tsujikawa, Shinji",
    title = "{Phantom-divide crossing and suppressed structure growth in kinetically braided dark energy with momentum exchange}",
    eprint = "2607.26447",
    archivePrefix = "arXiv",
    primaryClass = "astro-ph.CO",
    reportNumber = "WUCG-26-07",
    month = "7",
    year = "2026"
}

@article{Planck:2019nip,
    author = "Aghanim, N. and others",
    collaboration = "Planck",
    title = "{Planck 2018 results. V. CMB power spectra and likelihoods}",
    eprint = "1907.12875",
    archivePrefix = "arXiv",
    primaryClass = "astro-ph.CO",
    doi = "10.1051/0004-6361/201936386",
    journal = "Astron. Astrophys.",
    volume = "641",
    pages = "A5",
    year = "2020"
}

@article{Wang:2007shift,
    author = "Wang, Yun and Mukherjee, Pia",
    title = "{Observational Constraints on Dark Energy and Cosmic Curvature}",
    eprint = "astro-ph/0703780",
    archivePrefix = "arXiv",
    doi = "10.1103/PhysRevD.76.103533",
    journal = "Phys. Rev. D",
    volume = "76",
    pages = "103533",
    year = "2007"
}

@article{Chen:2018distance,
    author = "Chen, Lu and Huang, Qing-Guo and Wang, Ke",
    title = "{Distance Priors from Planck Final Release}",
    eprint = "1808.05724",
    archivePrefix = "arXiv",
    primaryClass = "astro-ph.CO",
    doi = "10.1088/1475-7516/2019/02/028",
    journal = "JCAP",
    volume = "02",
    pages = "028",
    year = "2019"
}

@article{Zhao:2010mg,
    author = "Zhao, Gong-Bo and Giannantonio, Tommaso and Pogosian, Levon and Silvestri, Alessandra and Bacon, David J. and Koyama, Kazuya and Nichol, Robert C. and Song, Yong-Seon",
    title = "{Probing modifications of General Relativity using current cosmological observations}",
    eprint = "1003.0001",
    archivePrefix = "arXiv",
    primaryClass = "astro-ph.CO",
    doi = "10.1103/PhysRevD.81.103510",
    journal = "Phys. Rev. D",
    volume = "81",
    pages = "103510",
    year = "2010"
}

@article{Song:2011mg,
    author = "Song, Yong-Seon and Zhao, Gong-Bo and Bacon, David and Koyama, Kazuya and Nichol, Robert C. and Pogosian, Levon",
    title = "{Complementarity of Weak Lensing and Peculiar Velocity Measurements in Testing General Relativity}",
    eprint = "1011.2106",
    archivePrefix = "arXiv",
    primaryClass = "astro-ph.CO",
    doi = "10.1103/PhysRevD.84.083523",
    journal = "Phys. Rev. D",
    volume = "84",
    pages = "083523",
    year = "2011"
}

\end{document}